\documentclass[onecolumn,showpacs,preprintnumbers,amsmath,amssymb]{revtex4}
\usepackage{graphicx} 
\usepackage{rotating}
\usepackage{bm}
\usepackage{bbm}
\usepackage{ulem} 
\usepackage[usenames]{color}

\newcommand{\be}{\begin{eqnarray}} 
\newcommand{\ee}{\end{eqnarray}}
\newcommand{\non}{\nonumber\\} 
\newcommand{\ave}[1]{\left\langle #1 \right\rangle}

\begin{document}

\preprint{APS/123-QED}

\title{Charge fluctuations and electric mass in a hot meson gas}
      
\author{M. D\"oring} \email{doering@ific.uv.es} \affiliation{
  Departamento de F\'{\i}sica Te\'orica and IFIC,
Centro Mixto Universidad de Valencia-CSIC,\\
Institutos de
Investigaci\'on de Paterna, Aptd. 22085, 46071 Valencia, Spain} 
\author{V. Koch}
  \email{vkoch@lbl.gov} \affiliation{Nuclear Science Division, Lawrence
  Berkeley National Laboratory, Berkeley 94720}

\date{\today}

\begin{abstract}
Net-Charge fluctuations in a hadron gas are studied using an effective
hadronic interaction. The emphasis of this work is to investigate the
corrections of hadronic interactions to the charge fluctuations of a
non-interacting resonance gas. Several methods, such as loop, density and virial
expansions are employed. The calculations are also extended to $SU(3)$ and some
resummation schemes are considered. Although the various corrections are
sizable individually, they cancel to a large extent. As a consequence we find
that charge fluctuations are rather well described by the free resonance gas.
\end{abstract}

\pacs{%
25.75.-q, 
12.38.Mh, 
25.75.Nq, 
11.10.Wx, 
24.60.-k  
}
\maketitle
\section{Introduction}

The study of event-by-event fluctuations or more generally fluctuations and
correlations in heavy ion collisions has recently received considerable
interest. Fluctuations of multiplicities and their ratios
\cite{mult_fluct}, transverse momentum
\cite{pt_fluct_na49,pt_fluct_ceres,pt_fluct_star,pt_fluct_phenix} and net charge
fluctuations \cite{star_q,na49_q,phenix_q,ceres_q} have
 been measured. Also first direct measurements of two
particle correlations have been carried out
\cite{trainor,phobos}. 

Conceptually, fluctuations may reveal
evidence 
of possible phase transitions and, more generally, 
provide information about the response functions of the system 
\cite{Jeon:2003gk}. For example, it is expected that near the QCD critical 
point long range correlation will reveal themselves  in enhanced
fluctuations of the transverse momentum ($p_t$) per particle
\cite{Stephanov:1999zu}.
Also, it has been shown
that the fluctuations of the net charge are sensitive to the fractional charges
of the quarks in the Quark Gluon Plasma (QGP)
\cite{Jeon:2000wg,Asakawa:2000wh}. 

Most fluctuation measures investigated so far are integrated ones, in the sense
that they are related to integrals of many particle distributions
\cite{koch_bialas}. Examples are: Multiplicity, charge and momentum
fluctuations which are all related to two-particle distributions. These
integrated measures have the advantage that they can be related to well
defined quantities in a thermal system. For example, fluctuations of the net
charge are directly related to the charge susceptibility. 
However, in an actual experiment additional, dynamical, i.e non-thermal
correlations may be present which make a direct
comparison with theory rather difficult. This
is particularly the case for fluctuations of the transverse momentum, where the
appearance of jet like structures provides nontrivial correlations
\cite{star_back2back,phenix_back2back,trainor}. These need to be understood and
eliminated from the analysis before fluctuation measurements can reveal
insight into the matter itself. 

In this article we will not be concerned with the comparison with experimental
data, and the difficulties associated with it. We rather want to investigate to
which extent interactions affect fluctuations. Specifically, we will study the
fluctuations of the net electric
charge of the system, the so-called charge fluctuations (CF). 
CF  have been proposed as a signature for the 
formation of the Quark Gluon Plasma (QGP) in heavy ion collisions
\cite{Asakawa:2000wh,Jeon:2000wg}. Refs.
\cite{Asakawa:2000wh,Jeon:2000wg} note that CF per degree
of freedom should be smaller in a QGP as compared  to a hadron gas
because the fractional charges of the quarks enter in square in the
CF. Using noninteracting hadrons and quarks, gluons,
respectively, it was found that the CF per entropy are about
a factor of 3 larger in a hadron gas than in a QGP. The net CF
per entropy has in the meantime been measured
\cite{star_q,na49_q,phenix_q,ceres_q}. At RHIC
energies the data are consistent with the expectations of a hadron gas, but
certainly not with that of a QGP. This might be due to limited acceptance as
discussed in \cite{bleicher,stephanov,gale}. 

The original estimates of the net charge
fluctuations per entropy in the hadron gas \cite{Asakawa:2000wh,Jeon:2000wg}
have been based on a system of noninteracting particles and resonances. While this
model has been proven very successful in describing the measured single
particle yields \cite{redlich,becattini}, it is not obvious to which extent
residual interactions among the hadronic states affect fluctuation 
observables. For example, in the QGP phase, lattice QCD calculations for the
charge
susceptibility and entropy-density differ from the result for a simple
weakly interacting QGP. Their ratio, however, agrees rather well with that of
a noninteracting classical gas of
quarks and gluons
\cite{Kapusta:1992fm,Jeon:2000wg,Jeon:2003gk,Allton:2005gk,Gupta_net_charge}.
As far as the hadronic phase is concerned, lattice results for charge
fluctuations are only available for systems with rather large pion
masses \cite{Allton:2005gk,Gupta_net_charge}. In this case, an appropriately
rescaled hadron gas model seems to describe the lattice results
reasonably
well \cite{redlich_lattice}. 
Lattice calculations with realistic pion masses, however, are not yet
available. Thus, one has to rely on hadronic model calculations in order to
assess the validity of the noninteracting hadron gas model for the description
of CF. In Ref. \cite{Eletsky:1993hv} the electric screening
mass $m_{\rm el}^2$ which is closely related to CF 
has been calculated up to next-to-leading (NLO) order in $\pi\pi$ interaction. However, the 
fact that thermal loops pick up energies
in the resonance region of the $\pi\pi$ amplitude where chiral perturbation theory is no longer 
valid leads to large theoretical uncertainties.

It is the purpose of this paper to provide a rough estimate of the effect of
interactions in the hadronic phase, in particular the effect of the coupling
of the $\rho$-meson to the pions. Since
$\rho$-mesons are
strong resonances which carry
the same quantum numbers as the CF this should
provide a good
estimate for the size of corrections to be expected from a complete
calculation; the latter will most likely come from lattice QCD, once numerically feasible.

As a first step we will consider the case of a heavy $\rho$-meson or,
correspondingly, a low temperature approximation. 
In this case the $\rho$-meson is not dynamical
and will not be part of the statistical ensemble. It will only induce an
interaction among the pions which closely corresponds to the interaction 
from the lowest order (LO) chiral Lagrangian. 

Although the temperatures in the hadronic phase are well below the $\rho$-mass,
it is interesting to estimate the residual $\pi\pi$ correlations introduced
when this resonance is treated dynamically. Special
attention is paid to charge conservation and unitarity. In addition, we will
investigate the importance of quantum statistics. Finally an extension to
strange degrees of freedom is provided.

This paper is organized as follows. After a brief review of the charge
fluctuations we introduce our model Lagrangian and discuss the heavy rho limit.
Next we discuss the treatment of dynamical $\rho$-mesons up to two-loop order and
compare with the results obtained in the heavy rho limit. Then, the effect of
quantum statistics and unitarity is discussed. Before we show
our final results including strange degrees of freedom, we will briefly comment
on possible resummation schemes.

\section{Charge fluctuations and Susceptibilities}
\label{sec:cf}
Before turning to the model interaction employed in this work, let us first
introduce some notation and recall the necessary formalism to calculate the
CF (for details, see, e.g., Ref. \cite{Jeon:2003gk}).

In this work we will consider a system in thermal equilibrium. In this case
the charge fluctuations $\ave{\delta Q^2}$ are given by the second derivative of the appropriate
free energy $F$ with respect to the charge chemical potential $\mu$: 
\be
\ave{\delta Q^2} = - T \frac{\partial^2 F}{\partial \mu^2} = -V T \chi_Q.
\label{charge_suscept}
\ee 
Here, $T$ ($V$) is the temperature (volume) of the system and $\chi_Q$ is the charge
susceptibility, which is often the preferred quantity to consider,
particularly in the context of lattice QCD calculations.  Equivalently, the
CF or susceptibility are related to the electromagnetic
current-current correlation function \cite{Ka1989,Prakash:2001xm} 
\be \Pi_{\mu
\nu}(\omega, {\bf k}) = \rm i \int dt d^3x \, e^{-i(\omega t - {\bf k x})}
\ave{J_\mu({\bf x},t) J_\nu(0)} \ee via \be \ave{\delta Q^2} = V T\,
\Pi_{00}(\omega=0,{\bf k}\rightarrow 0)=VT m_{{\rm el}}^2
\label{connection}
\ee 
which is illustrated for scalar QED in Appendix \ref{appendixa}.  Relation
(\ref{connection}) also establishes the connection between the CF and the electric screening mass $m_{\rm el}$.

As noted previously, the observable of interest is the ratio of CF over entropy 
\be D_S \equiv \frac{\ave{\delta Q^2}}{e^2 S}.
\label{observable}
\ee 
Given a model Lagrangian, both CF and entropy can be
evaluated using standard methods of thermal field theory (see
e.g. \cite{Ka1989}). CF are often evaluated via the
current-current correlator using thermal Feynman rules; evaluating the free
energy and using relation (\ref{charge_suscept}) will lead to
the same results as will be demonstrated in Sec. \ref{sec:rho_dynamical}.

Let us close this section by noting that in an actual experiment a direct
measurement of the entropy is rather difficult. However, the number of charged
particles $\langle N_{{\rm ch}}\rangle$ in the final state is a reasonable measure of the final state
entropy. Therefore, the ratio
\begin{equation}
D_c= 4 \; \frac{\langle\delta Q^2\rangle}{e^2\langle N_{{\rm ch}}\rangle}.
\label{mainobservable}
\end{equation}
has been proposed as a possible experimental observable for accessing the
CF per degree of freedom. For details and corrections to be
considered see Ref. \cite{Jeon:2003gk} and references therein.  In this
article we will concentrate on the ``theoretical'' observable $D_S$ defined in
Eq. (\ref{observable}).

\section{Model Lagrangian and $\boldsymbol{\pi\pi}$ interaction in the heavy $\boldsymbol{\rho}$ limit}
\label{sec:eff_interaction}
As already discussed in the Introduction, in this work we want to provide an estimate of the corrections to the CF introduced by
interactions among the hadrons in the hadronic phase. Since
it is impossible to
account for all hadrons and their interactions, we will concentrate on a system
of pions and $\rho$-mesons only, with some extensions to $SU(3)$ in later sections. A suitable effective Lagrangian for this
investigation is the ``hidden gauge'' approach of Refs. \cite{Bando:1984ej,Kaymakcalan:1984bz}. In this model  the $\rho$-meson is introduced 
as a massive gauge field. The $\pi\rho$ interaction results from the
covariant derivative
$D_\mu\Phi=\partial_\mu\Phi-\frac{ig}{2}\;\left[\rho_\mu,\Phi\right]$ acting
on the pion field $U(x)=\exp[i\Phi(x)/f_\pi]$ in the LO chiral Lagrangian 
\be {\cal L}_{\pi\pi}^{(2)}=\frac{f_\pi^2}{4}\;{\rm
Tr}\left[\partial_\mu U^\dagger \partial^\mu U+{\cal M}(U+U^\dagger)\right]
\label{chiral}
\ee
by the replacement $\partial_\mu \to D_\mu$.
Here,
\be
\Phi=\renewcommand{\arraystretch}{1.5} \left(\begin{array}{cc}
\pi^0&\sqrt{2}\pi^+\\ \sqrt{2}\pi^-&-\pi^0\end{array}\right),\quad
\rho_\mu=\renewcommand{\arraystretch}{1.5}\left(\begin{array}{cc}
\rho_\mu^0&\sqrt{2}\rho_\mu^+\\
\sqrt{2}\rho_\mu^-&-\rho_\mu^0\end{array}\right)
\label{su2_fields}
\ee
and $f_\pi=93$ MeV is the pion decay constant. 
An extension of the heavy gauge model to $SU(3)$ has been applied
for vacuum and in-medium processes (see,
e.g., Refs. \cite{Klingl:1996by,Alvarez-Ruso:2002ib}) and is straightforward
\cite{Marco:1999df}. 
This extension is considered in Sec. \ref{sec:su3extension}. 

The resulting $\pi\rho$ interaction terms are
\be 
{\cal L}_{\rho\pi\pi}
=\frac{ig}{4}\;{\rm Tr}\left(\rho_\mu \left[\partial^\mu\Phi,
\Phi\right]\right)
\label{34vertex}
\ee 
and 
\be 
{\cal L}_{\rho\rho\pi\pi}
=-\;\frac{g^2}{16}\;{\rm
Tr}\left(\left[\rho_\mu, \Phi\right]^2\right).
\label{4vertex}
\ee 
Chiral corrections to the interaction in Eq. (\ref{34vertex}) are of
${\cal O}(p^5)$ or higher as pointed out in Ref. \cite{Birse:1996hd}. The
interaction of Eq. (\ref{4vertex}) does not depend on the pion momentum, thus
violating the low energy theorem of chiral symmetry
\cite{Birse:1996hd}. Nevertheless, this term is required by the gauge
invariance of the $\rho$-meson \cite{Cabrera:2000dx} and in fact cancels
contributions in the pole term and crossed pole term of $\pi\rho$ scattering via
Eq. (\ref{34vertex}).

To leading order  in the pion field we, thus, have the following model
Lagrangian:
\be
{\cal L} =  {\cal L}_\Phi + {\cal L}_\rho + {\cal L}_{\rho\pi\pi} + 
 {\cal L}_{\rho\rho\pi\pi} 
\label{lagragian}
\ee
with the free field terms
\be
{\cal L}_\Phi &=& \frac{1}{4} {\rm Tr}\left(\partial_\mu \Phi \partial^\mu \Phi\right) 
- \frac{1}{4} {\rm Tr}\left(m_\pi^2 \Phi^2\right), \non
{\cal L}_\rho &=& -\frac{1}{8}\;{\rm Tr}\left(G_{\mu\nu}G^{\mu\nu}\right)
 + \frac{1}{4} {\rm Tr}\left(m_\rho^2 \rho_\mu\rho^\mu\right),
 \label{freefields}
\ee
and the interaction terms ${\cal L}_{\rho\pi\pi}$ and ${\cal
  L}_{\rho\rho\pi\pi}$ as given in Eq. (\ref{34vertex}) and (\ref{4vertex}),
respectively. For the kinetic tensor of the $\rho$,
$G_{\mu\nu}=\partial_\mu \rho_\nu-\partial_{\nu} \rho_{\mu}$, we restrict
ourselves to the Abelian part; a non-Abelian $\rho$ would lead to
additional $3\rho$ and $4\rho$ couplings. In the thermal loop expansion this would result in 
closed $\rho$ loops which are kinematically suppressed.
The coupling constant $g$ is fixed from the $\rho \rightarrow \pi
\pi$ decay to be $g = g_{\rho \pi \pi} = 6$ and and we use $m_\pi=138$ MeV 
and $m_\rho=770$ MeV throughout this paper.

As a first approximation, we start with the low temperature limit of the 
$\pi\pi$ interaction in which 
the $\rho$-meson mediates the interaction of the pions but does not enter 
the heatbath as an explicit degree of freedom.  
To this end we construct an effective interaction based on $s$-, $t$-, and $u$-channel $\rho$-meson
exchange as given by second order perturbation theory of  the interaction
${\cal L}_{\rho\pi\pi}$. Furthermore, we assume that the
momentum transfer $k^2$ of two pions interacting via a $\rho$ is much smaller
than the mass of the $\rho$--meson, $m_\rho^2>>k^2$, i.e. we replace the 
propagator of the exchanged $\rho$-meson
by $-1/m_\rho^2$. Thus, we arrive at the following effective interaction
\be {\cal L}^{{\rm eff}}_{\pi\pi}&=&\frac{g^2}{2m_\rho^2}\big((\pi^-
\stackrel{\leftrightarrow}{\partial_\mu}\pi^+)^2 -2(\pi^0
\stackrel{\leftrightarrow}{\partial_\mu} \pi^+)(\pi^0
\stackrel{\leftrightarrow}{\partial^\mu}\pi^-)\big).
\label{eff_strong}
\ee
Note that in this limit, subsequently referred to as the 
``heavy $\rho$ limit'' the $\rho\rho\pi\pi$ term from
Eq. (\ref{4vertex}) does not contribute at order $g^2$. 

The effective Lagrangian of Eq. (\ref{eff_strong}) shows the identical 
isospin and momentum structure as the kinetic term of Eq. (\ref{chiral})
at $1/f_\pi^2$. However, comparing the overall coefficient one arrives at
\be 
m_\rho^2=3f_\pi^2g^2
\label{ksfr}
\ee 
which differs by a factor of $3/2$ from the well known KSFR  relation
\cite{KSFR} $m_\rho^2 = 2 f_\pi^2 g^2$. 
We should point out the same factor 
has been observed in the context of the anomalous 
$\gamma\pi\pi\pi$ interaction \cite{Cohen:1989es}. As discussed in more detail
in Appendix \ref{appendixb} the KSFR relation is recovered if one restricts
the model to the $s$-channel diagrams for the isovector $p$-wave ($T_{11}$) 
amplitude. Once also $t$- and $u$-channels are taken into account the factor $3/2$
appears. 
For this study, we prefer the interaction (\ref{eff_strong}) over 
${\cal L}_{\pi\pi}^{(2)}$ from Eq. (\ref{chiral}) as it delivers a better data 
description at low energies in the $\rho$-channel (see Appendix \ref{appendixb}). 
The simplification from the ''heavy $\rho$'' limit of the $\pi\pi$
interaction will later be relaxed in favor of dynamical $\rho$-exchange.
However, the interaction in the heavy $\rho$ limit will still serve as a
benchmark for the more complex calculations.

Since we are interested in the electromagnetic  
polarization tensor, the interaction of
Eq. (\ref{eff_strong}), together with the kinetic term of the pion, is gauged
with the photon field by minimal substitution, leading to 
\be {\cal
L}_{\pi\gamma}&=&-\frac{1}{4}\left(F^{\mu\nu}\right)^2-m_\pi^2\big(\pi^+\pi^-
+\frac{1}{2}(\pi^{0})^2\big)+(D_\mu^* \pi^-)(D^\mu \pi^+)
+\frac{1}{2}\:\left(\partial_\mu \pi^0\right)^2, \nonumber \\
&+&\frac{g^2}{2m_\rho^2} \left(\pi^-D_\mu \pi^+ -\pi^+ D_\mu^* \pi^-
\right)^2-\frac{g^2}{m_\rho^2}\left(\pi^0 D^\mu \pi^+-\pi^+ \partial^\mu
\pi^0\right) \left( \pi^0 D_\mu^* \pi^--\pi^-\partial_\mu \pi^0\right)
\label{effL}
\ee 
with the covariant derivative of the photon field 
$D_\mu=\partial_\mu+ie A_\mu$, $e>0$, and the photon field tensor
$F^{\mu\nu}$, leading to the $\gamma\pi\pi$ and 
$\gamma\gamma\pi\pi$ interactions of scalar QED,
plus $\gamma\pi\pi\pi\pi$
and $\gamma\gamma\pi\pi\pi\pi$ vertices. 
Vector meson dominance leads to
$\gamma\rho^0$ mixing as pointed out, e.g., in Ref. 
\cite{Klingl:1996by}, additionally to the vertices of
Eq. (\ref{effL}). However, since the correlator of Eq. (\ref{connection}) is
evaluated at the photon point, the form factor is unity and the 
 process $\gamma\to\rho^0\to\pi\pi$ which emerges in the systematic approach of
Ref. \cite{Schechter:1986vs} does
not contribute to the $\gamma\pi\pi$ coupling. Thus, no modification of
Eq. (\ref{effL}) is required. Note also that the anomalous interaction
providing $\gamma\pi\rho$ vertices \cite{Klingl:1996by} does
not contribute in the long-wavelength limit studied here. This follows a
general rule noted in Ref. \cite{Kapusta:1992fm}.

In the following chapters, the $\rho$ will be also treated dynamically. The 
interaction with the photon is then given by the scalar QED vertices from above, 
plus a $\gamma\rho\pi\pi$ vertex which is obtained from Eq. (\ref{34vertex})
by minimal substitution. 
With the same procedure the direct $\gamma\rho$ interaction 
is constructed
from Eq. (\ref{freefields}), 
leading to the vertices 
\be
\includegraphics[width=2.5cm]{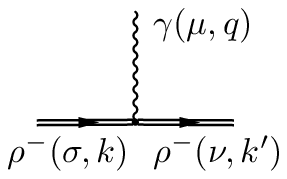} \;\;{\hat=}\;e\left(k^\nu
g_{\mu\sigma}+k'^\sigma g_{\mu\nu}-(k+k')^\mu g_{\sigma\nu}\right),
\quad
\includegraphics[width=2cm]{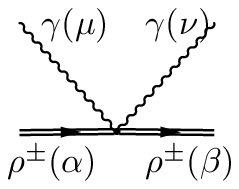}\;\;{\hat
=}\;2e^2\left(g_{\mu\beta}g_{\alpha\nu}-g_{\mu\nu}g_{\alpha\beta}\right) 
\label{gamma_rho}
\ee 
in the imaginary time formalism.

\section{Charge fluctuations at low temperatures}
\label{sec:charge-fluct-contact}
Having introduced the effective interaction in the heavy $\rho$ limit, we can
evaluate the correction to the CF due to
this interaction. Before discussing the results 
let us first remind the reader
about the basic relations for CF in a noninteracting gas of pions and $\rho$-mesons.

\subsection{Charge fluctuations  for free pions and $\rho$-mesons}
\label{sec:free_pion_gas}
\begin{figure}
\includegraphics[width=6.5cm]{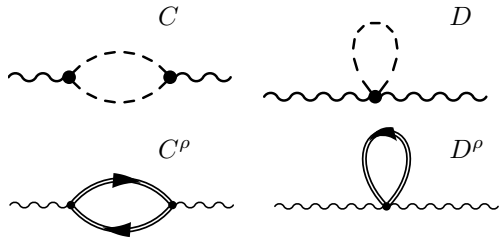}
\caption{Photon selfenergy at $e^2$ for the free pion gas ($C,\,D$) and the free $\rho$ gas ($C^\rho,\,D^\rho$).}
\label{fig:simplebubbles}
\end{figure}

In order to illustrate the relations of Sec. \ref{sec:cf} and to establish
a baseline it is instructive to calculate $D_S$ from 
Eq. (\ref{observable}) for the
free pion gas in two ways: once via
Eq. (\ref{connection}) and also directly from statistical mechanics.  The
interaction from Eq. (\ref{effL}) reduces to scalar QED in the zeroth order in
$g$. To order $e^2$ the selfenergy is given by the set of gauge invariant
diagrams in Fig. \ref{fig:simplebubbles}
and reads  \be  \Pi^{00}(k^0=0, {\bf k\to 0})=e^2\left(C+D\right),\quad
\ave{\delta Q^2} =e^2 T V \left(C+D\right)
\label{self_e_square}
\ee 
according to Eq. (\ref{connection}) with 
\be
C=\frac{1}{\pi^2}\int\limits_0^\infty dp\;\omega \; n[\omega],\quad
D=\frac{1}{\pi^2}\int\limits_0^\infty dp\;p^2\;\frac{n[\omega]}{\omega}
\label{c_d}
\ee 
where $\omega=\sqrt{p^2+m_\pi^2}$ the pion energy,
$n[\omega]=1/(\exp\left(\beta\omega\right)-1)$ the Bose-Einstein factor, and
$\beta=1/T$.  The CF from
Eq. (\ref{self_e_square}) can also be derived from  statistical mechanics, 
\be
\ave{\delta Q^2}=e^2 T^2\;\frac{\partial^2}{\partial\mu^2}|_{\mu=0}\log Z,
\label{stat_mech_cf}
\ee 
\be \log Z_0(\mu)=-V\int\frac{d^3p}{\left(2\pi\right)^3}\;\sum_{\mu_i=\pm\mu,
0} \log\left(1-e^{-\beta(\omega+\mu_i)}\right).
\label{stat_mech_cf_2}
\ee 
Both the photon selfenergy and Eq. (\ref{stat_mech_cf})
lead to the same CF also at the perturbative level as will be
seen in Sec. \ref{sec:rho_dynamical}.
The value for the chemical potential of $\mu_i=\pm \mu$ in Eq. (\ref{stat_mech_cf_2}) corresponds
to charged pions and $\mu_i=0$ is assigned to neutral pions which do not
contribute to the CF but  to the  entropy 
$S=\partial(T \log Z)/\partial T$ of the free gas, 
\be
S_0=\frac{1}{2\pi^2}\;\frac{V}{T}\;\int\limits_0^{\infty} dp\;
p^2\;n[\omega]\left(3\omega+\frac{p^2}{\omega}\right).
\label{free_entropy}
\ee 
In the high temperature limit, or for massless pions, the relevant thermodynamical
quantities are given by 
\be
\ave{\delta Q^2}=\frac{e^2V}{3} T^3, \quad
S=\frac{2\pi^2V}{15} T^3,\quad \langle N_{\rm ch}\rangle=\frac{2\zeta(3)V}
{\pi^2} T^3
\label{t_limit}
\ee 
where $\langle N_{\rm ch}\rangle$ is defined as in
Ref. \cite{Jeon:2000wg}. For the quantity $D_S$ from
Eq. (\ref{observable}) we obtain $D_S=0.185$ for massive free pions at
$T=170$ MeV and $D_S=0.253$ for massless pions. 
For $D_c$ from Eq. (\ref{mainobservable}), 
the values are $4.52$ and $5.47$, respectively.

The classical (Boltzmann) limit is obtained by  replacing 
the Bose-Einstein distribution
$n$ in Eqs. (\ref{c_d}) and (\ref{free_entropy}) by the Boltzmann distribution 
$n_{\rm B}=\exp(-\beta\omega)$. In this case at $T=170$ MeV we obtain $D_S=0.156$ and
$D_S=1/6$ for massive and massless pions, respectively. 
For all masses and temperatures, $D_c=4$ the classical limit.
For a QGP made out of massless quarks and gluons, $D_S = 0.034$, following
the same arguments as in \cite{Jeon:2000wg}. This is about a factor
of five smaller than a pion gas.

The CF for the free $\rho$ gas are given by the diagrams with the double lines in Fig. \ref{fig:simplebubbles}.
With the $\rho$ propagator 
\be
D^{\mu\nu}=\frac{1}{k^2-m_\rho^2+i\epsilon}\left(g^{\mu\nu}-\frac{k^\mu
k^\nu}{m_\rho^2}\right)
\label{rho_propagator}
\ee 
and the interaction from Eq. (\ref{gamma_rho}) the photon selfenergy turns out to be 
\be \Pi_{\rho}^{00}(k^0=0, {\bf k\to 0})=3 e^2 [C^\rho+D^\rho]
\label{free_rho_gas}
\ee 
where the upper index means that the pion mass is substituted by the
$\rho$-mass in $C$ and $D$ from Eq. (\ref{c_d}). The factor of three
corresponds to the sum over the physical polarizations of the $\rho$. The same
factor also appears in $\log Z_0$ of Eq. (\ref{stat_mech_cf_2}) for the
$\rho$.

\subsection{$\boldsymbol{\pi\pi}$ interaction in the heavy $\boldsymbol{\rho}$ limit to order $\boldsymbol{e^2g^2}$}
\label{sec:effective_results}
At order $e^2g^2$ the Feynman rules derived from the heavy $\rho$ limit Eq. (\ref{effL}) 
lead to the set of five diagrams (eff1)
to (eff5) depicted in Fig. \ref{fig:diagrams}. They are gauge invariant as shown in Appendix \ref{appendixc3}.
\begin{figure}
\includegraphics[width=9cm]{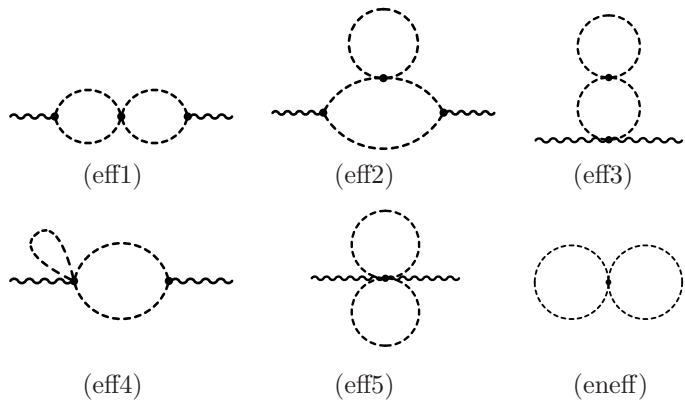}
\caption{Selfenergy for $\pi\pi$ 
interaction in the heavy $\rho$ limit at order $e^2g^2$ (diagrams (eff1) to (eff5)). Expansion of
$\log Z$ at $g^2$ for the calculation of the entropy (diagram (eneff)).}
\label{fig:diagrams}
\end{figure}
The summation over Matsubara frequencies has been performed by a
transformation into contour integrals following Ref. \cite{Ka1989}. The limit
$\displaystyle{(k^0=0,{\bf k}\rightarrow{\bf 0})}$ for the external photon has to be taken before
summation and integration, as discussed in Appendix \ref{appendixa}. The loop momenta
factorize so that the diagrams of
Fig. \ref{fig:diagrams} can be expressed in terms of the quantities $C$ and
$D$ from Eq. (\ref{c_d}) as shown in Tab. \ref{tab:num_res_eff}.
\begin{table}
\caption{Static selfenergy $\Pi^{00}(k^0=0, {\bf k\to 0})$ from Fig. \ref{fig:diagrams} with $C$ and $D$ from Eq. (\ref{c_d}).}
\begin{tabular*}{0.37\textwidth}{@{\extracolsep{\fill}}ll}
\hline\hline Diagram& Contribution \\ \hline
\rule[-4mm]{0mm}{10mm}(eff1)&$\displaystyle{-\frac{3}{2}\frac{e^2g^2}{m_\rho^2}\;C^2}$
\\
\rule[-4mm]{0mm}{9mm}(eff2)&$\displaystyle{-\frac{e^2g^2}{m_\rho^2}\;D\big(D-3C-\beta\;\frac{\partial}{\partial\beta}\;(C-D)\big)}$
\\
\rule[-4mm]{0mm}{9mm}(eff3)&$\displaystyle{+\frac{e^2g^2}{m_\rho^2}\;D(2D-C)}$
\\ \rule[-4mm]{0mm}{9mm}(eff4)&$\displaystyle{-5\frac{e^2g^2}{m_\rho^2}\;CD}$
\\
\rule[-4mm]{0mm}{9mm}(eff5)&$\displaystyle{-\frac{5}{2}\frac{e^2g^2}{m_\rho^2}\;D^2}$
\\ \hline\hline
\end{tabular*}
\label{tab:num_res_eff}
\end{table}
The sum of the diagrams is cast in a surprisingly simple form,
\begin{equation}
\sum_{i=1}^5 \Pi_i(k^0=0, {\bf k}\to{\bf
0})=-\frac{e^2g^2}{m_\rho^2}\left[\frac{3}{2}\left(C+D\right)^2 -\beta
\;D\;\frac{\partial}{\partial \beta}\left(C-D\right)\right].
\label{pinullsum}
\end{equation}
The entropy correction at $g^2$ is calculated from $\log Z$ given by diagram
(eneff) in Fig. \ref{fig:diagrams}, \be
S_1=-\frac{3g^2V}{2T}\;\left(\frac{m_\pi}{m_\rho}\right)^2\;D\left(C+D\right).
\label{s_1}
\ee 
Note that using the LO chiral Lagrangian from Eq. (\ref{chiral})
instead of the $\pi\pi$ interaction in the heavy $\rho$ limit, results would
simply change by a factor of
$(2/3)^2$, up to tiny corrections, which are due to higher order contributions
involving the chiral symmetry breaking term  $\sim{\cal M}$ from Eq.
(\ref{chiral}). Numerical results can
be found
in Sec. \ref{sec:numres1}, which supersede our findings from Ref.
\cite{Doring:2002qa}. 

\section{The $\boldsymbol{\rho}$--meson in the heatbath}
\label{sec:rho_dynamical}
In this section we will relax the assumption of a heavy non-dynamical
$\rho$-meson. 
 This will allow
for an estimate of the CF from the residual interactions of the
$\rho$ when this particle is treated as an explicit degree of freedom.
It will also avoid some problems induced in the calculation from vertices of
higher order in momenta such as encountered in
the ${\cal L}_{\pi\pi}^{(4)}$ calculation
in Ref. \cite{Eletsky:1993hv}
(see discussion in Sec. \ref{why}, \ref{sec:numres1}).
The $\rho\pi\pi$ interaction from Eq. (\ref{34vertex}) involves vertices only
linear in momentum and a smoother temperature
dependence is expected.

\begin{figure}
\includegraphics[width=10cm]{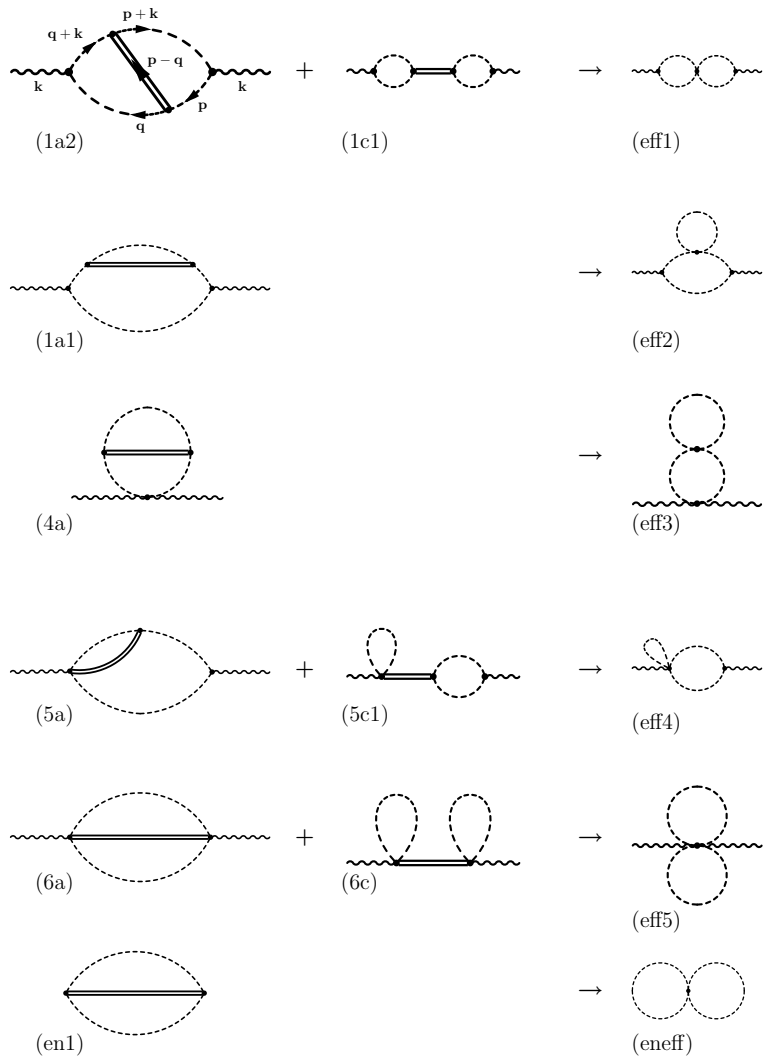}
\caption{Overview of the relevant two-loop diagrams at $e^2g^2$ for the 
photon selfenergy and at $g^2$ for the entropy. Diagram (1a2) is 
calculated in detail in Appendix \ref{appendixcc}, where also the results for all other diagrams and a proof of gauge invariance are found.
The diagrams on the right hand side (eff1-eff5) correspond to
the heavy $\rho$ limit of the ones given on the left. This limit, indicated with arrows, is numerically shown
in Appendix \ref{appendixc}. }
\label{fig:overview_2loop}
\end{figure}

We start with the calculation of the diagrams in the first two
columns of Fig. \ref{fig:overview_2loop}
because this subset 
corresponds to the heavy $\rho$ limit from Sec. \ref{sec:eff_interaction};
by increasing the $\rho$-mass from its physical value to infinity
in these diagrams, the previous results from 
Tab. \ref{tab:num_res_eff} are recovered as illustrated in Appendix \ref{appendixc}.
Note that there is no need to include $\gamma\rho^0$ mixing or anomalous vertices as we have already seen in Sec. \ref{sec:eff_interaction}. 

Here and in the following sections, the $\rho$ is treated as a stable particle (propagator from Eq. (\ref{rho_propagator}))
and we ignore imaginary parts at the cost of unitarity violations as will be discussed
in Sec. \ref{why}. A $\rho$ with finite width would induce problems concerning
gauge invariance:
one would have to couple the photon to all intermediate $\rho$ selfenergy diagrams 
that build up the $\rho$ width in the Dyson-Schwinger summation.
In principle, this is possible --- see the last part of Sec. \ref{sec:numres} --- but goes beyond the scope of this work.
The results for the diagrams with dynamical $\rho$ from Fig. \ref{fig:overview_2loop} are found in Eq. (\ref{analytic_2_loop},\ref{easyloops}) and
Fig. \ref{fig:num2l} of Appendix \ref{appendixc}, together with a detailed calculation of
one of the diagrams and a discussion of the infrared divergences. In Appendix \ref{appendixc3} the gauge invariance of the diagrams is shown.

\begin{figure}
\includegraphics[width=10cm]{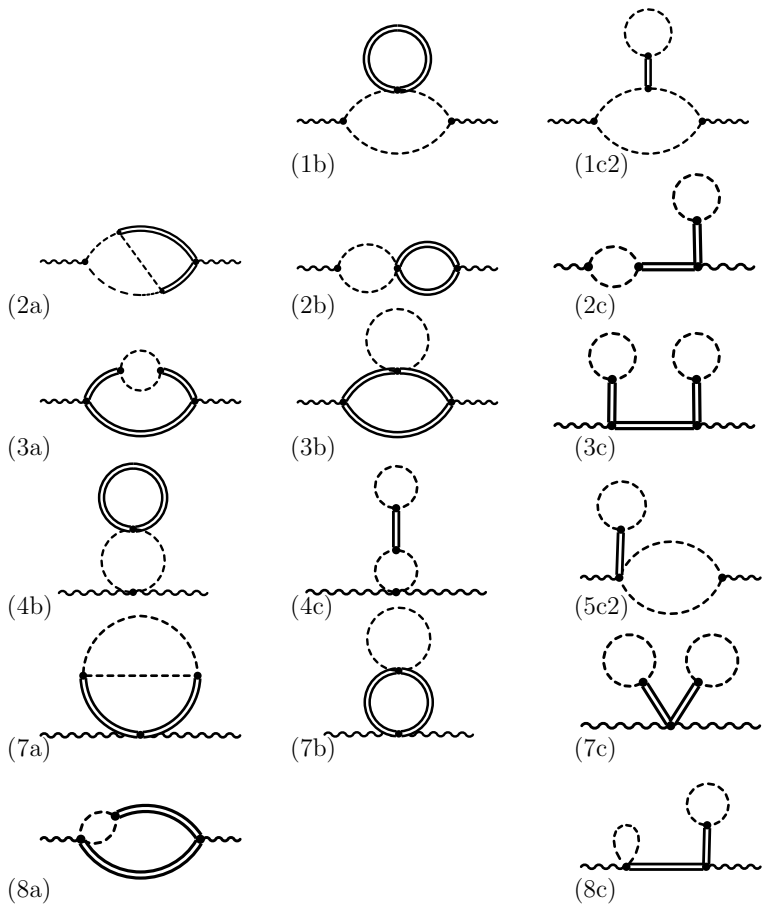}
\caption{Additional, sub-leading, diagrams at $e^2g^2$ with direct $(\gamma)\gamma\rho\rho$ couplings and with $\rho\rho\pi\pi$ interaction. Also, the diagrams which vanish are shown [(1c2), (2c), (3c), (4c), (5c2), (7c), (8c)].}
\label{fig:overview_more}
\end{figure}
At order $e^2g^2$ there are additional diagrams with direct $\gamma\rho\rho$ and 
$\gamma\gamma\rho\rho$ couplings from Eq. (\ref{gamma_rho}) and also with the $\rho\rho\pi\pi$ coupling from Eq. (\ref{4vertex}) which is required by the gauge invariance of the $\rho$-meson. The resulting diagrams are displayed in Fig. \ref{fig:overview_more}. Some of these diagrams contain more than one $\rho$-propagator. They are sub-dominant because every $\rho$ propagator counts as $1/m_\rho^2$. Furthermore, Fig. \ref{fig:overview_more} shows diagrams
which have a closed pion loop with only one vertex of the $\rho\pi\pi$ type (see, e.g., diagram (2c)). The latter diagrams vanish due to the odd integrand in the loop integration. The set of diagrams from Figs. \ref{fig:overview_2loop} and \ref{fig:overview_more} is complete at order $e^2g^2$.

\begin{figure}
\includegraphics[width=13cm]{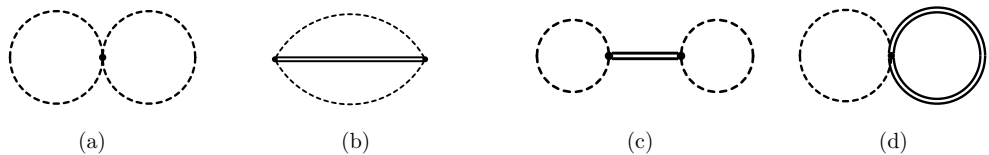}
\caption{Correction to $\log Z(\mu)$. Diagram (a) shows the 
$\pi\pi$ interaction in the heavy $\rho$ limit, diagrams (b)--(d) the interaction via 
explicit vector meson from Eqs. (\ref{34vertex})
and (\ref{4vertex}).}
\label{fig:entropies}
\end{figure}

The non-vanishing diagrams from Fig.
\ref{fig:overview_more} are best calculated by evaluating the corresponding
partition function,  $\log Z$, at finite chemical potential $\mu$ and
differentiating with respect to $\mu$ \cite{Kapusta:1992fm,Eletsky:1993hv} (see
also Eq. (\ref{stat_mech_cf})). For a calculation at finite $\mu$ we first convince ourselves that for the
simple interaction from Eq. (\ref{eff_strong})
the use of Eq. (\ref{stat_mech_cf}) leads to the same 
results as in Sec. \ref{sec:effective_results}.
The calculation at finite $\mu$ implies a shift in the zero-momenta of the
propagators and derivative vertices, $p^0\to p^0\pm\mu$
\cite{Eletsky:1993hv,Ka1989}, depending on the charge states of the
particles. The correction to $\log Z(\mu)$ from diagram (a) in
Fig. \ref{fig:entropies} 
with the interaction from Eq. (\ref{eff_strong}) is given by 
\be \log Z_{{\rm (a)}}(\mu)=\frac{-g^2}{16\;
m_\rho^2}\;\beta
V\left[3\left(V_+-V_-\right)^2+m_\pi^2\left(U_++U_-\right)\left(4D+U_++U_-\right)\right]
\label{logmu_eff}
\ee 
with 
\be U_\pm=\frac{1}{\pi^2}\int\limits_0^\infty
dk\;\frac{k^2}{\omega}\;n[w\pm \mu],\quad V_\pm=\frac{1}{\pi^2}
\int\limits_0^\infty dk\;k^2\;n[w\pm \mu]
\label{u_v}
\ee 
and $D$ from Eq. (\ref{c_d}). Applying Eq. (\ref{stat_mech_cf}) to $\log
Z_{{\rm (a)}}(\mu)$ reproduces the result for the photon selfenergy in the
heavy $\rho$ limit from Eq. (\ref{pinullsum}) which is shown to be gauge
invariant in Appendix \ref{appendixc3}.

Thus having established that equivalence of photon selfenergy and charge
fluctuations (Eq. (\ref{connection})) holds on the perturbative level, we are 
encouraged to evaluate the diagrams of  Fig. \ref{fig:overview_more} by
differentiating the appropriate terms in $\log Z$ with respect to the chemical
potential.  
The diagrams for $\log Z$ corresponding to the photon self energies given in
Figs. \ref{fig:overview_2loop} and \ref{fig:overview_more} are displayed in Fig.
\ref{fig:entropies}(b,c,d).
(Details can be found in Appendix \ref{appendixc2}).

In Fig. \ref{fig:compare_rhos} corrections to the electric mass
 of a free pion gas due to
different sets of diagrams are shown. As a reference, we also plot the
results for gases of
noninteracting pions and noninteracting $\rho$-mesons (''free $\pi$'' and ''free $\rho$'').
\begin{figure}
\includegraphics[width=9.5cm]{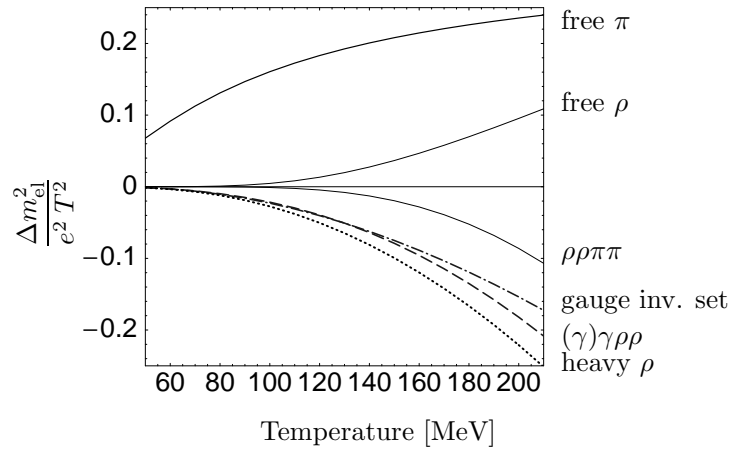} 
\caption{Corrections to $m_{\rm el}$ or CF. Dashed-dotted line:
result 
from the gauge invariant subset 
of diagrams from the first two columns of Fig. \ref{fig:overview_2loop}.
Dashed line: Result from diagrams (b)+(c) from 
Fig. \ref{fig:entropies}.  Dotted 
line: heavy $\rho$ limit from Sec. \ref{sec:effective_results}. 
Solid lines: free $\pi$ gas, free $\rho$ gas, and the $\rho\rho\pi\pi$ 
interaction from Fig. \ref{fig:entropies} (d).}
\label{fig:compare_rhos}
\end{figure}
The electric mass from the diagrams of Fig. \ref{fig:diagrams} with the $\pi\pi$ interaction in the heavy $\rho$ limit is plotted as the dotted line. The electric mass from the diagrams in the first two columns of Fig. \ref{fig:overview_2loop} with dynamical $\rho$ is plotted as the dashed-dotted line. 
At low temperatures, both results coincide (in detail this is also plotted in Fig. \ref{fig:num2l}). However, at higher temperatures we observe significant differences which shows, thus, that the $\rho$ obtains importance as an explicit degree of freedom.

The diagrams (b) and (c) from Fig. \ref{fig:entropies} correspond to
the first two columns of Fig. \ref{fig:overview_2loop}. Additionally, they provide photon selfenergies with $\gamma\rho\rho$ and
$\gamma\gamma\rho\rho$ vertices from Fig. \ref{fig:overview_more}, diagrams (2a), (3a), (7a), and (8a). 
As shown in Fig. \ref{fig:compare_rhos} (dashed line), these
additional $\gamma\rho$ couplings obtain some minor influence above $T\sim
150$ MeV.

Additionally, in Fig. \ref{fig:overview_more} there are diagrams with
$\rho\rho\pi\pi$ couplings from Eq. (\ref{4vertex}). The diagrams (1b), (2b), (3b), (4b), and (7b) correspond
to diagram (d) in Fig. \ref{fig:entropies}. In the heavy $\rho$ limit these diagrams do not contribute. However, for dynamical $\rho$-mesons these diagrams contribute significantly due to the sum over the spin of the $\rho$. In Fig. \ref{fig:compare_rhos} the
resulting electric mass is displayed as the solid line (''$\rho\rho\pi\pi$'').

\section{Relativistic virial expansion}
\label{sec:relavir}

In Ref. \cite{Eletsky:1993hv} the electric mass has been determined using chiral $\pi\pi$ interaction and thermal loops leading to 
results that show 
large discrepancies to a virial calculation of $m_{\rm el}^2$. Before we
discuss these differences in Sec. \ref{why}, \ref{sec:numres1} let us review the theoretical
framework first. 
The virial expansion is an expansion of thermodynamic quantities in
powers of the fugacities $e^{\beta\mu}$, while the interaction enters as
experimentally measured phase-shifts. Consequently, all
orders of the interaction are taken into account. Thermal
loops, on the other hand, respect quantum statistics (Bose-Einstein in our case)
and, thus, contain an infinite subclass of the virial expansion. However, the
interaction only enters up to a given order. Thus, the loop and virial expansion
represent quite different approximations and it will depend on the problem at
hand which is the more appropriate one. The effect on quantum statistics can be
considerable. For example at $T=170$ MeV the values for the electric mass $m_{\rm el}^2$ of the free $\pi$ gas or the two-loop diagrams, Eq.
(\ref{pinullsum}), change by 20\% and 38\% (!) respectively, if we take the
Boltzmann limit. Therefore, it is desirable to
have a density expansion that respects particle statistics as well as
sums all orders of the interaction. While this might be very difficult if not
impossible to do in general, it can be done up to second order in the
(Bose-Einstein) density.

The partition function can separated into a free and an interacting part,
\be
\log Z=\log Z_0+\sum_{i_1,i_2}z_1^{i_1}\,z_2^{i_2}\,b(i_1,i_2)
\ee
in an expansion in terms of the chemical potential $\mu$ with $z_j=\exp(\beta\mu_j)$ for $j=1,2$ the fugacities. 
In the $S$-matrix formulation of statistical mechanics from Ref. \cite{Dashen}
the second virial coefficient $b(i_1,i_2)$ can be separated into a statistical
part and a kinematic part containing the vacuum $S$-matrix according to
\be
b(i_1,i_2)=\frac{V}{4\pi i}\int\frac{d^3 {\bf k}}{(2\pi)^3}\int
dE\;e^{-\beta\sqrt{{\bf k}^2+E^2}} \;{\rm Tr}_{i_1,i_2}
\left[AS^{-1}(E)\;\frac{\stackrel{\leftrightarrow}\partial}{\partial E}\;S(E)\right]_c
\label{virialb2}
\ee
where $A$ is the (anti)symmetrization operator for interacting (fermions) bosons
and the trace is over the sum of connected diagrams (index ''$c$''). In Eq.
(\ref{virialb2}), $V$ is the Volume, $k$ is the momentum of the $n$-particle cluster
in the gas rest frame and $E=s^{1/2}$ stands for the total c.m. energy. The labels $i_1,i_2$ indicate a channel of the $S$-matrix with
$i_1+i_2$ particles in the initial state. For the second virial coefficient,
$i_1=i_2=1$. 

For $\pi\pi$ scattering, Eq. (\ref{virialb2}) can be integrated over $k$ and the
$S$-matrix can be expressed via phase shifts,
weighted with their degeneracy \cite{Eletsky:1993hv}. With $B_2=b(i_1,i_2)/V$ in the limit $V\to\infty$ one obtains
\be
B_2^{(\pi\pi),\;{\rm
Boltz}}(\mu=0)&=&
\frac{1}{2\pi^3 \beta}\;\int\limits_{2m_\pi}^\infty
dE\;E^2\;K_2(\beta E)\sum_{\ell,
I}(2I+1)(2\ell+1)\;\frac{\partial\delta_\ell^I(E)}{\partial E}
\non
&=&
\frac{1}{2\pi^3}\;\int\limits_{2m_\pi}^\infty
dE\;E^2\;K_1(\beta E)\sum_{\ell, I}(2I+1)(2\ell+1)\;\delta_\ell^I
\label{b2known}
\ee
where the second line has been obtained after integration by parts (assuming
$\delta_\ell^I \rightarrow 0$ as $E\rightarrow 2 m_\pi$).
The sum over phase shifts $\delta_\ell^I$ (isospin $I$, angular momentum
$\ell$) is restricted to $\ell+I=$ even and $K_i$ are the modified Bessel
functions of the second kind.
The virial expansion in this or similar form has been applied in 
numerous studies of the thermal properties of interacting hadrons as, e.g., \cite{Welke:1990za,Venugopalan:1992hy}, among them the electric mass 
\cite{Eletsky:1993hv}. Note that, e.g. in \cite{Eletsky:1993hv}, Bose-Einstein statistics is taken into account for the non-interacting, free gas, part.
However, it is also possible to include particle statistics for the interacting part. This means the summation of
so-called exchange diagrams as outlined in Ref. \cite{Dashen}, Sec. VIIB. We employ this idea and also include a finite
chemical potential. This is achieved by projecting the binary collisions of
pions in different charge states to the isospin channels \cite{Eletsky:1993hv}. Additionally, the interaction $T$ matrix is boosted from the gas rest frame to the two-particle c.m. frame \footnote{Note this boost to the two-particle c.m. frame is merely a convenience as the scattering amplitude is easily obtained in this frame. The boost is not essential. For a detailed discussion see Appendix \ref{app:deriva}.} and the $T$-matrix is defined via phase shifts with the final result
\be
B_2^{(\pi\pi),\;{\rm Bose}}(\mu)&=&\frac{\beta}{4\pi^3}\;\int\limits_{2m_\pi}^\infty dE\int\limits_{-1}^1 dx\int\limits_0^\infty dk\;\frac{E\;k^2}{\sqrt{E^2+k^2}}\bigg[\delta_0^2(E)\left(n[\omega_1+\mu]n[\omega_2+\mu]+n[\omega_1-\mu]n[\omega_2-\mu]\right)\non
&+&\delta_0^2(E)\left(n[\omega_1+\mu]n[\omega_2]+n[\omega_1-\mu]n[\omega_2]\right)
+3\;\delta_1^1(E)\left(n[\omega_1+\mu]n[\omega_2]+n[\omega_1-\mu]n[\omega_2]\right)\non
&+&\delta_0^2(E)\left(\frac{1}{3}n[\omega_1+\mu]n[\omega_2-\mu]+\frac{2}{3}n[\omega_1]n[\omega_2]\right)+
3\;\delta_1^1(E)\;n[\omega_1+\mu]n[\omega_2-\mu]\non
&+&\delta_0^0(E)\left(\frac{2}{3}n[\omega_1+\mu]n[\omega_2-\mu]+\frac{1}{3}n[\omega_1]n[\omega_2]\right)\bigg].
\label{b2mu}
\ee
A more explicit derivation of this result can be found in Appendix \ref{app:deriva}.
The first line of Eq. (\ref{b2mu}) corresponds to $\pi\pi$ scattering with a net
charge of the $\pi\pi$ pair of $|C|=2$, the second line to $|C|=1$ and
the third and 4th line to
$C=0$. 
The boosted Bose-Einstein factors which arise after summations over exchange diagrams are
\be
&&n[\omega_{1,2}\pm \mu]=\frac{1}{e^{\beta(\omega_{1,2}\pm \mu)}-1},\quad
\omega_1=\gamma_f\left(\frac{1}{2}\;E+\frac{k\;Q\;x}{\sqrt{E^2+k^2}}\right),\quad
\omega_2=\gamma_f\left(\frac{1}{2}\;E-\frac{k\;Q\;x}{\sqrt{E^2+k^2}}\right),\non
&&\gamma_f=\left(1-\frac{k^2}{E^2+k^2}\right)^{-\frac{1}{2}},\quad Q=\frac{1}{2}\sqrt{E^2-4m_\pi^2}
\label{boostbose}
\ee
with the momentum of the pion $Q\equiv Q_{{\rm c.m.}}$ in the two-pion c.m. frame. 

Obviously, the chemical potential can not be factorized in Eq. (\ref{b2mu}) so that the expansion is rather in 
powers of Bose-Einstein factors $n$ than in powers of
$e^{\beta\mu}$ as in a conventional virial expansion. Eq.
(\ref{b2mu})  
contributes also to higher virial coefficients. The situation resembles the case
of a free Bose-Einstein gas that contributes to all virial coefficients which
can be seen by expanding the Bose-Einstein factor in powers of $e^{\beta \mu}$.
Therefore, in the following we will refer to the expansion
(\ref{b2mu}) as ``(low) density expansion``. The term ''virial expansion'' will
be reserved for the well known expansion in terms of classical (Boltzmann)
distributions. We note, that 
in the Boltzmann limit the standard expression for the virial coefficient, e.g.
Eq. (9) of Ref. \cite{Eletsky:1993hv}  is recovered; setting additionally $\mu=0$ we obtain Eq. (\ref{b2known}).

The connection of $B_2(\mu)$ to physics is given by
\be
\log Z(\mu)=VB_2(\mu),\quad P(\mu)=\frac{B_2(\mu)}{\beta},\quad m_{\rm el}^2=e^2\left(\frac{\partial^2 P}{\partial \mu^2}\right)_{\mu=0}
\label{melvir}
\ee
where $P$ is the correction to the pressure 
Note that for the electric mass the contribution $\sim \delta_0^0$ vanishes 
in the Boltzmann limit (and is small anyways).
The form of Eq. (\ref{b2mu}) makes it as easy to use as the common virial
expansion, inserting the $\pi\pi$ phase shifts $\delta_0^0$, $\delta_1^1$, and
$\delta_0^2$ which we adopt from Ref. \cite{Welke:1990za}. 
The inelasticities of the $\pi\pi$ amplitude are small in the relevant energy
region and we have not taken them into account in Eq. (\ref{b2mu}). 

\subsection{Density expansion versus thermal loops}
\label{why}
It is instructive to see to which extent the thermal loop expansion and the extension of the virial
expansion from Eq. (\ref{b2mu}) agree. To this end we need to match both approaches by extracting
the scattering amplitude from our model Lagrangian and insert it into Eq.
(\ref{b2mu}). For simplicity, we first study the $\pi\pi$ interaction in the heavy
$\rho$ limit at $g^2$ and evaluate Eq. (\ref{b2mu}).
As this interaction is not unitary, one has to go back to the original
$S$-matrix formulation and express it in terms of the (on-shell) $T$-matrix
\cite{Dashen} which can then be calculated from theory. 
Given the normalization
of the $T$-matrix used in this paper,
$S=1-\frac{iQ}{8\pi\sqrt{s}}\;T$, the right hand side of Eq. (\ref{virialb2})
can be written as
\be
\left(S^{-1}\;\frac{\partial S}{\partial E}-\frac{\partial S^{-1}}{\partial E}\;S\right)=
-\frac{i}{8\pi}\frac{\partial}{\partial E}\left[\frac{Q}{E}\left(T+T^\dagger\right)\right]
+\frac{1}{64\pi^2}\left(\frac{Q}{E}\;T^\dagger \right)\frac{\stackrel{\leftrightarrow}\partial}{\partial E}\;
\left(\frac{Q}{E}\;T \right).
\label{ttos}
\ee 
Using the relation between $S$-matrix and phase shifts, $S=e^{2i\delta}$, we
find
\be
\frac{\partial}{\partial E}\, \delta_\ell^I\;\hat{=}\;
-\;\frac{\partial}{\partial E}\left(\frac{2\,Q}{E}\;{\rm Re}\,T_\ell^I\right)+ \frac{8\,Q^2}{E^2}\left({\rm Re}\,T_\ell^I \;\frac{\stackrel{\leftrightarrow}\partial}{\partial E}\;{\rm Im}\,T_\ell^I\right)
\label{deltatcorresp}
\ee
where the connection between isospin amplitudes $T^I$ and their projection into
partial waves $T^I_\ell$ is given in Eq. (\ref{pwaintegral}). 
Inserting this expression into Eq. (\ref{b2mu}) leads to the density expansion
based on a given model amplitude.
We note that the second term in Eq. (\ref{deltatcorresp}) is quadratic in the
amplitude and vanishes for real amplitudes. Therefore, close to threshold,
where the amplitudes are small and real, the quadratic term can be neglected.
However, with increasing energy unitarity requires that the  imaginary
part of the amplitude will become sizable so that the second term cannot
any longer be neglected. This is  especially the case if the amplitude
is resonant. Consequently, the use of point-like interactions at tree level
which are always real and not unitary
might lead to rather unreliable predictions for thermodynamic
quantities. 

Before we discuss the importance of unitarity, let us first establish that the
density expansion of Eq. (\ref{b2mu}) and the loop expansion lead to the same
results if both methods are based on the same point-like interaction.
The partial amplitudes $T^I_\ell$ for the $\pi\pi$ interaction in the heavy $\rho$ limit are obtained from Eq. (\ref{t1_rho}) by neglecting $s,t,u,$ and $\Gamma$ in the denominators and $s\equiv E^2$. 
Inserting the result in Eq. (\ref{b2mu}) and calculating the pressure from Eq.
(\ref{melvir}) we obtain exactly the same result as for the thermal loops from
Eq. (\ref{logmu_eff}) at $\mu=0$. We have also verified that this agreement
holds in a simple  $\phi^4$ theory of uncharged interacting bosons.
Calculating the electric mass in both approaches 
for the $\pi\pi$ interaction in the heavy $\rho$ limit (Eqs. (\ref{pinullsum},\ref{connection}) and (\ref{b2mu},\ref{melvir})), we again find
perfect agreement. 

Consequently, and not so surprisingly, both thermal loop and
density expansion lead to the same result, if the interaction in the density
expansion is truncated at the appropriate (unitarity violating) level. This is
also true in the classical (Boltzmann) limit. In this limit, a similar equivalence has been
found in
\cite{Eletsky:1993hv} using an effective range expansion for the amplitude;
see also \cite{Bugrii:1995yg} for a related equivalence for propagators.

While it is comforting to see that both approaches agree in the same order of
density and interaction, this agreement highlights a possible problem for the
loop expansion. If the order of the interaction considered violates unitarity
the second term of Eq. (\ref{deltatcorresp}) is ignored and the loop expansion may
lead to unreliable results for the pressure etc. This is of particular
importance if the amplitudes are resonant, as it is the case for the $\rho$-exchange. 

\begin{figure}
\includegraphics[width=8.5cm]{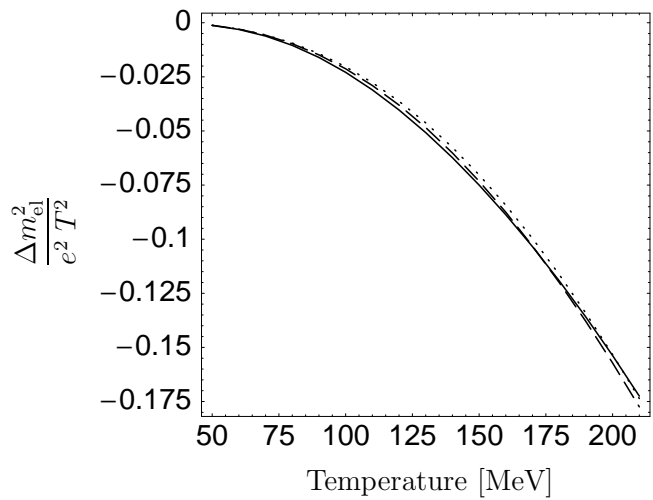}
\caption{Electric mass from dynamical $\rho$ exchange. Solid line: from the diagrams in the first two columns of Fig. \ref{fig:overview_2loop}. Dotted line: Bose-Einstein density expansion from dynamical $\rho$ exchange (no imaginary parts, $\Gamma_\rho\to 0$). Dashed line: Same, but $\Gamma_\rho=150$ MeV.}
\label{fig:compadynrho}
\end{figure}

In order to see these effects we concentrate on the gauge invariant set of
diagrams given in the first two columns of Fig. \ref{fig:overview_2loop}. The
result for these diagrams is given in Eqs. (\ref{analytic_2_loop},
\ref{easyloops}) and plotted in Fig. \ref{fig:compadynrho} as the solid line. 
In the calculation of these thermal loops we have made the following
approximations, see Appendix \ref{appendixc}: (I) The poles of the $\rho$ have
been neglected in the contour integration (see the explanation following
Eq. (\ref{trickder})).  (II) The $\rho$ has no width, i.e. the $\rho$ propagator
is given by $D^{\mu\nu}$ from Eq. (\ref{rho_propagator}). (III) Only the real
parts of the thermal loops have been considered. 

In the following we test these approximations by comparing the thermal loop
result with a suitable ``toy model'' low density expansion. 
For the interaction driving the low density expansion we take the partial waves
from Eq. (\ref{t1_rho}) and project out the $T_\ell^I$ by the use of Eq.
(\ref{pwaintegral}). Furthermore, we set $\Gamma_\rho=0$ in Eq. (\ref{t1_rho})
in this interaction. Third, we consider only the term linear in $T$ in Eq.
(\ref{deltatcorresp}) for the density expansion. This means that imaginary parts
are neglected. The low density expansion, constructed in this way, exhibits the
same approximations (II) and (III) as the calculation of the thermal loops
above, i.e. the zero width and the reduction to the real part only. The result
of this ``toy model'' low density expansion is plotted in Fig.
\ref{fig:compadynrho} as the dotted line.

Both the results from thermal loops (solid line) and the density expansion
(dotted line) agree closely. The small deviation of both curves is due to the
additional approximation (I) which we have made in the calculation of the
thermal loops, i.e. neglecting the poles in the contour integration. Note also
that other partial waves than $T_0^0$, $T_1^1$, and $T_0^2$ are present in the
results from the thermal loops because the $\rho$ exchange contains all partial
waves. However, from the agreement found here, we may conclude that these higher
partial waves give negligible contributions (at least in the present
$\rho$-exchange model).

In our ``toy model'' low density expansion, we can allow for a finite width in
the $\rho$-propagator. This implies that the 
$\rho$-propagator is given by $D_\rho=[p^2-m_\rho^2+im_\rho\Gamma(\sqrt{s})]^{-1}$, where
$\Gamma(\sqrt{s}) = \Gamma(m_\rho)(m_\rho^2/E^2)(E^2-4m_\pi^2)^{3/2}
/(m_\rho^2-4m_\pi^2)^{3/2}$. With this modification, we evaluate again the
electric mass. However, as we still use only the term linear in $T$ in Eq.
(\ref{deltatcorresp}), any imaginary parts of the amplitude arising from the
finite width are still ignored.  
The result is shown as the dashed line in Fig. \ref{fig:compadynrho}; the
electric mass hardly changes.  

\begin{figure}
\includegraphics[width=9.5cm]{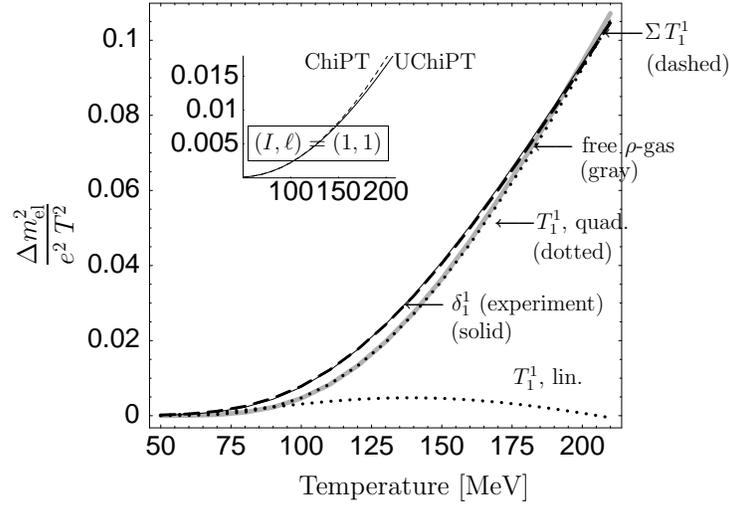}
\caption{{\it Main plot:} Contribution to $m_{\rm el}^2$ in the $(I,\ell)=(1,1)$-channel of $\pi\pi$ scattering. 
Solid line: From experimental phase shift. Dotted lines: Terms linear and quadratic in $T$ from Eq. (\ref{mel_rhoex}) with $T$
from a unitary $\rho$ exchange model. Dashed line: Sum of linear and quadratic term. Gray line: free $\rho$-gas (Boltzmann). \newline{\it Insert:} With $T$ from the LO chiral Lagrangian. Dashed line: Tree level. Solid line: Unitarization with $K$-matrix.}
\label{fig:tests}
\end{figure}

Let us now discuss the effect of the imaginary parts of the amplitude. To
simplify the discussion let us restrict ourselves to the vector-isovector
$(I,J)=(1,1)$ channel, which is dominated by the $\rho$-resonance. We will also
work in the Boltzmann limit as effects due to unitarity are
independent of the statistical ensemble. The model amplitude is simply the
$s$-channel $\rho$-exchange diagram with a $\rho$-propagator as given above.
This amplitude is unitary by construction and describes the scattering  data in
the $(I,J)=(1,1)$-channel well (see  Fig. \ref{fig:pipivac}). 
With the (complex) $T$-matrix $T_1^1$ the electric mass $m_{\rm el}$ is given by
\be
m_{\rm el}^2(\mu=0,I=\ell=1)&=&-\frac{6\,e^2\,\beta}{\pi^3}\int\limits_{2m_\pi}^\infty dE\;Q\,E\,K_1(\beta E)\;{\rm Re}\,T_1^1+
\frac{24\,e^2}{\pi^3}\int\limits_{2m_\pi}^\infty dE\;Q^2\;K_2(\beta E)\;\big({\rm Re}\,T_1^1\;\frac{\stackrel{\leftrightarrow}\partial}{\partial E}\;{\rm Im}\,T_1^1\big)\non
\label{mel_rhoex}
\ee 
where the first term is linear and the second quadratic in the amplitude. It
is the second, quadratic, term where the imaginary part of the amplitude
enters. In Fig. \ref{fig:tests} the different contributions to the electric mass
according to the decomposition Eq. (\ref{mel_rhoex}) are plotted. 
As a reference we also show the result using experimentally measured phase
shift $\delta_1^1$ (solid black line).  
Obviously the contribution from the quadratic term (``$T_1^1$, quad.'') is
dominant, and adding the linear (``$T_1^1$, lin.'') and quadratic
terms we obtain good agreement with the result from the $\delta_1^1$
experimental phase shift. This is to be expected as $T_1^1$ fits vacuum data
well. Note that the linear term alone vastly underpredicts the electric mass.
Thus the imaginary part of the amplitude is essential for the proper description
of the fluctuations. 

Furthermore, the electric mass from the free $\rho$-gas
(Boltzmann statistics) (gray line, Fig. \ref{fig:tests}) agrees well with the
results from the experimental phase shift and unitary $\rho$-model. Indeed, it
can be shown that the $\pi\pi$-interaction via unitary $s$-channel
$\rho$-exchange in the limit of vanishing width leads to a contribution to 
$\log Z$ equal to that of a free $\rho$-gas \cite{Welke:1990za,Dashen:1974jw}.
In this limit $\delta_1^1(E)=\pi\,\Theta(E-m_\rho)$ allowing for an explicit
evaluation of Eq. (\ref{b2mu}) in the Boltzmann approximation. For Bose-Einstein
statistics the situation is more complicated. In Ref. \cite{Dashen:1974jw} it
has been shown for meson-baryon interaction that also in this case the
interaction of two particles via a narrow resonance $N^*$ leads to the same
grand canonical potential as from a free $N^*$-gas with the corresponding
Fermi-statistics for $N^*$; however, the proof requires a self-consistent medium
modification of the $N^*$ width and a consideration of larger classes of
diagrams.

While in our toy model we could simply restore unitarity by introducing
a $\rho$ width, in a more complete calculation this is considerably more
difficult. For example, using a $\rho$-propagator with a width in the
diagrams of Figs. \ref{fig:overview_2loop} and \ref{fig:overview_more} leads to
additional photon couplings to the intermediate pion loops, which generate the
$\rho$-width. This is simply a consequence of gauge invariance (see
Appendix \ref{appendixc3}). Therefore, introducing unitary amplitudes while
maintaining gauge invariance is a non-trivial task.

An alternative approach to assess the role of unitarity is to unitarize a given
amplitude using the $K$-matrix approach (see, e.g. \cite{Weise}). This
approach does not add any additional dynamics, and therefore provides a good
estimator on the importance of unitarity alone. Using the $K$-matrix approach we
can in principle take any of the interactions discussed in this paper. Here we
choose the interaction in the $(1,1)$-channel from the LO chiral Lagrangian
given in Eq. (\ref{chiral}). Details of the calculation can be found in Appendix
\ref{appendixb1}. Maintaining gauge invariance in a $K$-matrix unitarization
scheme requires special care and is beyond the scope of this paper. Ignoring
this issue, we can compare the electric mass from the unitarized version using
Eqs. (\ref{unit},\ref{b2mu},\ref{melvir}) with the tree level amplitude using
$T_1^1$ from Eq. (\ref{tree_level_l2}) and then Eqs.
(\ref{deltatcorresp},\ref{b2mu},\ref{melvir}). The results are plotted in the
insert of Fig. \ref{fig:tests} and show only a small correction due
to unitarization. 

Consequently, unitarity by itself is not as crucial
as the dynamics which generates the resonance. In other words as long as the
phase shift is slowly varying with energy unitarity corrections are small. A
resonant amplitude on the other hand corresponds to a very rapidly varying
phase-shift. Since it is the derivative of the phase-shift which enters the
density expansion, resonant amplitudes are expected to dominate. Consequently,
a resonance gas should provide a good leading order description of the
thermodynamics of a strongly interacting system.

Note that the unitarized amplitude $T_{(u),1}^1$ from Eq. (\ref{tree_level_l2})
corresponds to a unitarization via the Bethe-Salpeter equation in the limit
where the real
parts of the intermediate $\pi\pi$-loops are neglected; the freedom in the
choice of the real part (loop regularization) can be used to fit to experimental
phase shifts, which in turn introduces the missing dynamics (see, e.g.,
Refs. \cite{Oller:2000ma,Doring:2004kt}) This should lead to more reliable
predictions \cite{in_preparation}.

To conclude this analysis of the density expansion, it appears that the low
density expansion of Eq. (\ref{b2mu}),
using experimental phase shifts, will give the most reliable 
results, while a simple hadron gas calculation should provide a
reasonable first estimate for the fluctuations of a system. 
Finally, there are certain features of the $\rho$ model from Sec.
\ref{sec:rho_dynamical} which can not be taken into account in the low density
expansion: The $\rho\rho\pi\pi$ and $(\gamma)\gamma\rho\rho$ interactions
discussed in Sec. \ref{sec:rho_dynamical} (Fig. \ref{fig:overview_more})
are a consequence of the $\rho$ being treated as a heavy gauge particle; these
features will be missed in the low density or virial expansion in which
the $\rho$ is not more than a resonant structure in the $\pi\pi$ amplitude.
These considerations will be taken into account in the final numerical result
from Sec. \ref{sec:numres}.

\subsection{Numerical results for the interacting pion gas}
\label{sec:numres1}
\begin{figure}
\includegraphics[width=8.5cm]{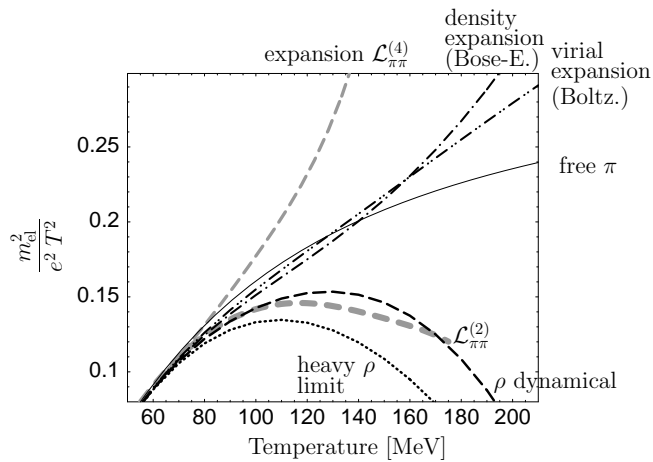}
\caption{CF or electric mass for the interacting 
$\pi\pi$ system. The ${\cal L}_{\pi\pi}^{(2)}$, ${\cal L}_{\pi\pi}^{(4)}$ calculations are from Ref. \cite{Eletsky:1993hv}
and the conventional virial expansion (Boltzmann statistics) reproduces results from Ref. \cite{Eletsky:1993hv}.}
\label{fig:compare_first_kapusta}
\end{figure}

In Fig. \ref{fig:compare_first_kapusta} the results so far obtained are
compared 
to Ref.  \cite{Eletsky:1993hv} (gray dashed lines). 
The electric mass for pions interacting in the heavy $\rho$ limit from 
Sec. \ref{sec:effective_results} is indicated with the dotted line. 
Taking into account that the interaction 
from Eq. (\ref{eff_strong}) is around 3/2 times stronger than the one 
from the LO chiral Lagrangian, the calculation is consistent
with the 
${\cal L}_{\pi\pi}^{(2)}$ calculation from Ref. \cite{Eletsky:1993hv} which we
have also checked analytically. 
The result for dynamical $\rho$ exchange 
(black dashed line) contains the contributions from free pion and 
$\rho$ gas and the diagrams from Fig. \ref{fig:entropies} (b), (c), 
and (d). The difference to the heavy $\rho$ limit shows the importance of the $\rho$ as an explicit degree
of freedom in the heatbath. 

Up to $T\sim 130$ MeV the dynamical $\rho$ exchange contributes with the same sign as the virial expansion
from Ref. \cite{Eletsky:1993hv} although they differ largely in size
due to the lack of imaginary part in the loop calculation, especially in
the $(I,\ell)=(1,1)$-channel as discussed above. Also the $\rho$-model
does not describe the $(I,\ell)=(2,0)$ amplitude very well. 

For the low density expansion from Eq. (\ref{b2mu}) and the virial expansion
from Eq. (\ref{b2known}) we use the phase shifts from Ref. \cite{Welke:1990za}.
Note that there is a partial cancellation from the $\delta_1^1$ and $\delta_0^2$ partial waves
\cite{Eletsky:1993hv}.

The ${\cal L}_{\pi\pi}^{(4)}$ calculation from Ref. \cite{Eletsky:1993hv},
which of course contains also the ${\cal L}_{\pi\pi}^{(2)}$ contribution, 
shows a very distinct result. The reason is twofold: on one hand, unitarity is
not preserved (see discussion in Sec. \ref{why}). On the other hand, the thermal
loops in the ${\cal L}_{\pi\pi}^{(4)}$ calculation pick up high
c.m. momenta where the theory
is no longer valid and the dependence of the NLO interaction on high
powers of momenta introduces artifacts. Note that the size of the correction from
${\cal L}_{\pi\pi}^{(4)}$ alone is larger than the one from ${\cal
L}_{\pi\pi}^{(2)}$ for $T>80$ Mev.

\begin{figure}
\includegraphics[width=8.5cm]{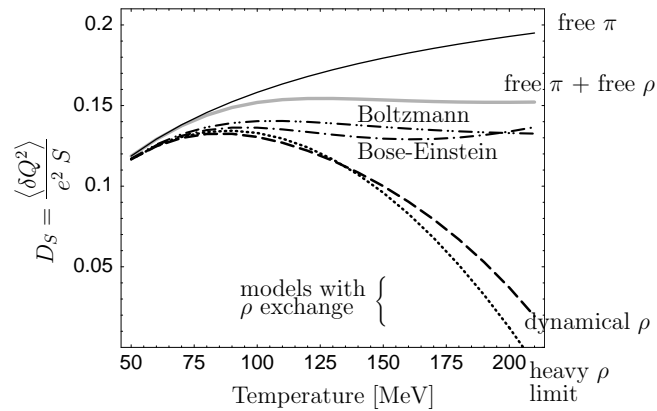}
\caption{$D_S$ from Eq. (\ref{observable}) for the interacting $\pi\pi$ system. The heavy $\rho$ limit from Eqs. (\ref{pinullsum},\ref{s_1}) includes the contribution from the free $\pi$ gas. The contributions from free $\pi$ and free $\rho$ gases are included for the result from Eqs. (\ref{logb},\ref{logc},\ref{logzeight}), the latter indicated as ''dynamical $\rho$''. The virial and density expansions are indicated according to their underlying particle statistics with ''Boltzmann'' and ''Bose-Einstein'', respectively.
}
\label{first_dqds}
\end{figure}

The results for the observable $D_S$ from Eq. (\ref{observable}) are displayed in
Fig. \ref{first_dqds}. Corrections to the entropy are included: from
Eq. (\ref{s_1}) for the heavy $\rho$ limit, from
Eq. (\ref{analytic_2_loop}) for the case with dynamical $\rho$ and from Eq. (\ref{melvir}) for the low density expansions.
For the thermal loops, indicated by ''models with $\rho$ exchange'', $D_S$ is
suppressed. This 
is due to the large negative correction to $\ave{\delta Q^2}$ as has been seen in
Fig. \ref{fig:compare_first_kapusta}. The virial expansion and the density
expansion coincide better with each other than in Fig.
\ref{fig:compare_first_kapusta} and can be roughly approximated by a gas of
noninteracting pions and rhos. 

Having contrasted virial expansions and dynamic $\rho$ model in Sec. \ref{why},
the most realistic results for CF and $D_S$ for the interacting $\pi\pi$ system
are given by the Bose-Einstein density expansion from Eq. (\ref{b2mu}). While
this result is somewhat below the estimate of a gas of free pions and
$\rho$-meson, it is  nowhere near the value of $D_S^{QGP} \simeq 0.034$ for the
quark gluon plasma. 
 
\section{Higher order corrections}
\label{sec:necklace}
Both the density expansions and $\rho$ models from the last sections are quadratic in density, i.e., the statistical factor $n$.
However, at the temperatures of the hadronic phase higher effects in density play an important role. Virial expansions become complicated beyond
the second virial coefficient and no experimental information exists on three body correlations.
Performing resummations is, therefore, of interest. This will include the density and strong coupling $g$ to all orders. Of course, this can not be  done in a systematic way; resummations only contain certain classes of diagrams at a given order in perturbation theory. In all resummations, $\log Z$ is calculated at finite $\mu$ and then Eq. (\ref{stat_mech_cf}) is applied in order to obtain the electric mass. We have convinced ourselves in Sec. \ref{sec:rho_dynamical} that this is a charge conserving procedure.

\begin{figure}
\includegraphics[width=8cm]{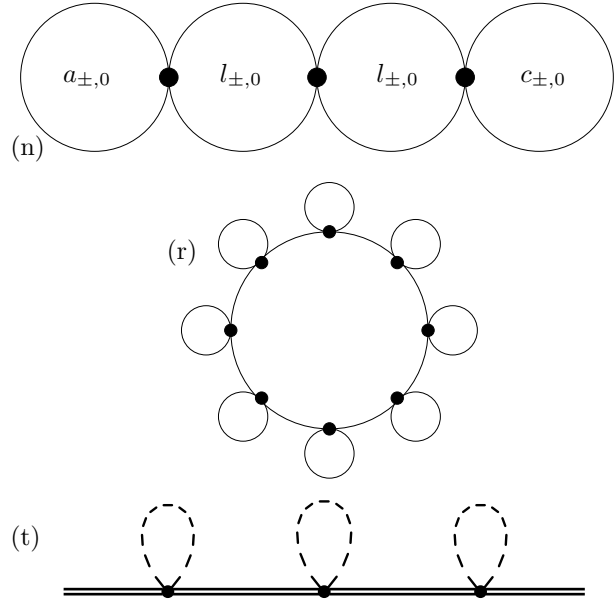}
\caption{Resummation schemes: necklace (n) and ring (r). Below, the tadpole medium correction of the $\rho$ propagator is displayed (t).}
\label{fig:resum_schemes}
\end{figure}

We start with two natural extensions of the basic interaction diagram (a) in 
Fig. \ref{fig:entropies}, displayed in Fig. \ref{fig:resum_schemes} (n) and (r) 
using for both of them the effective interaction
of the heavy $\rho$ limit from Eq. (\ref{eff_strong}). Alternatively, one can use the LO chiral Lagrangian from Eq. (\ref{chiral}). As found in Sec. \ref{sec:eff_interaction}, results for the dominant part of this interaction are obtained by simply multiplying $g$ in the following by a factor of $2/3$. However, one should keep in mind the unitarity problems of these simplified point-like interactions which have been addressed in Sec. \ref{why}.

\begin{figure}
\includegraphics[width=7.5cm]{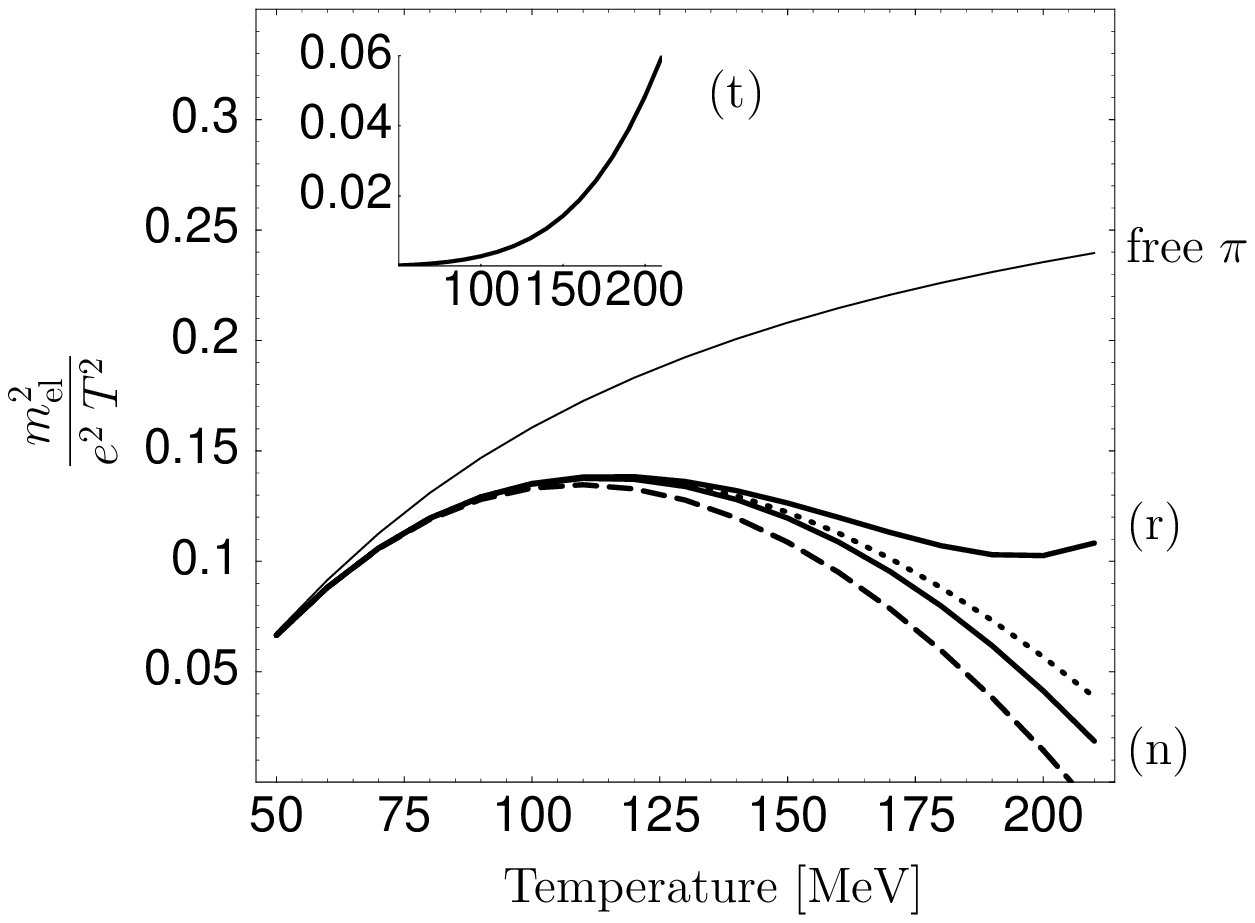}\hspace*{0.5cm}
\includegraphics[width=8.5cm]{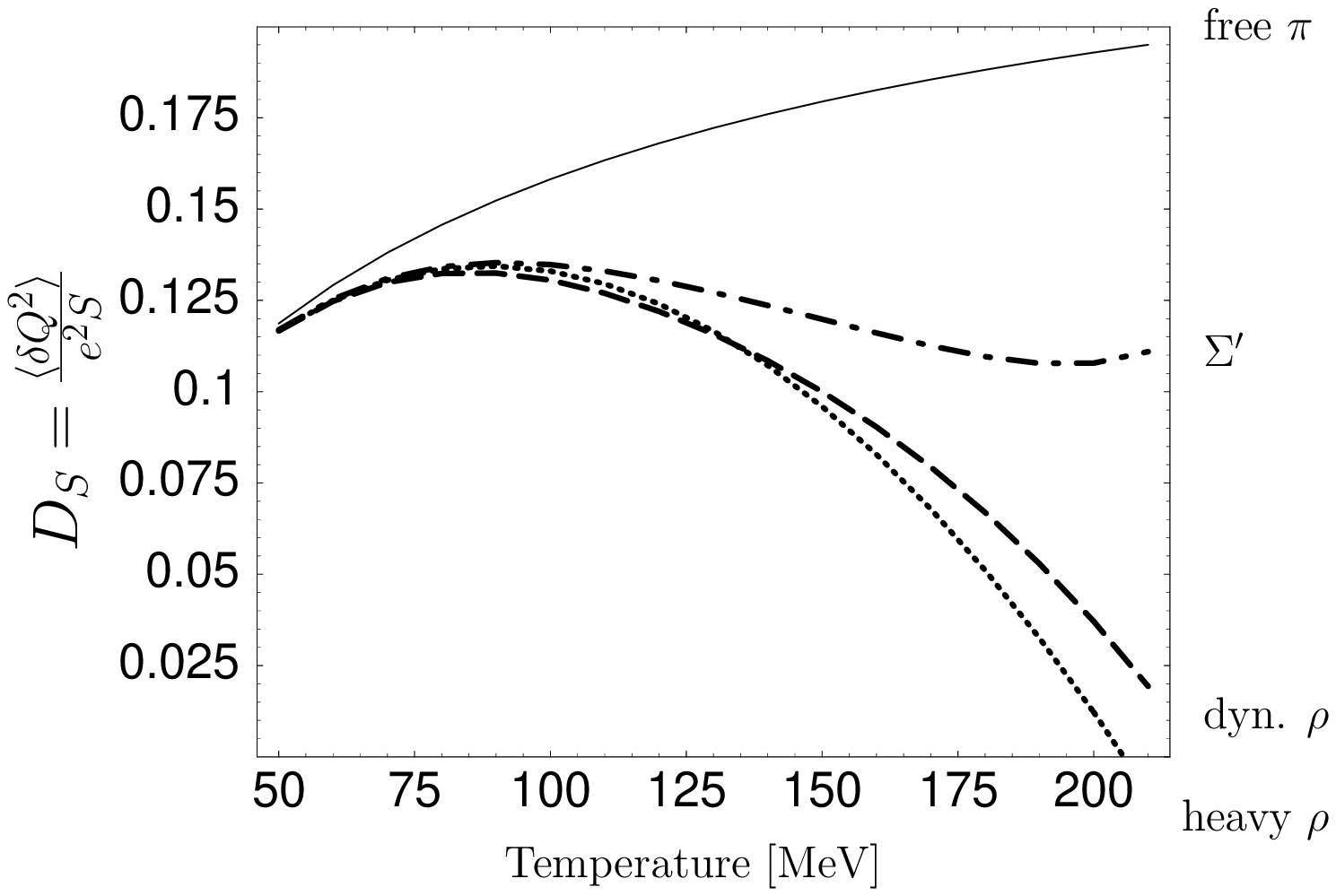}
\caption{{\it \underline{Left panel}, main plot:} Resummations (n) and (r) from
Fig. \ref{fig:resum_schemes} and their expansions up to $g^2/m_\rho^2$ (dashed
line)
and 
up to $g^4/m_\rho^4$ (dotted line) which are the same for (n) and (r). {\it Insert:} resummation (t), for order
$g^4/m_\rho^4$ and higher.
\underline{{\it Right panel:}} CF over entropy, $D_S$. Result for
heavy $\rho$ limit and dynamical $\rho$ as in Fig. \ref{first_dqds}. The result including 
the resummations (see text) is shown as the dashed
dotted line ($\Sigma'$).}
\label{fig:resummed}
\end{figure}

For the calculation of diagram (n) we utilize an equation
of the Faddeev type. The Faddeev equations, usually used in three-body scattering processes
as in Ref. \cite{Doring:2004kt} in a different context, are an easy way to sum
processes whose elementary building blocks are of different types, as in
this case loops of neutral pions with chemical potential $\mu=0$ and charged
loops: 
\be \log Z_{({\rm n})}(\mu)&=&\frac{1}{2}\;\beta
V\left(\frac{1}{2}\;a_0\;b_\pm+a_\pm\left(b_\pm+b_0\right)\right)\non
b_\pm&=&g'c_\pm+g'l_\pm\left(b_\pm+b_0\right)\non
b_0&=&\frac{1}{2}\;g'c_0+\frac{1}{2}\;g'l_0\;b_\pm
\label{Faddeev}
\ee with $g'=-g^2/m_\rho^2$. The first loop in the chain is labeled $a$, the
last one $c$, and $l$ means an intermediate loop. 
The indices ''$\pm$'' and ''$0$''
label charged and uncharged loops, respectively. It is instructive to expand
Eq. (\ref{Faddeev}) loop by loop which shows that the structure indeed
reproduces all sequences of charged and uncharged loops, of all lengths. There
is a symmetry factor of $1/2$ for every loop of neutral pions and a global
factor of $1/2$ for every pion chain.
The solution of Eq. (\ref{Faddeev}) is found in Appendix \ref{appendixd}. The result of resummation (n) 
is plotted in Fig. \ref{fig:resummed} together with its
expansion up to $g^2/m_\rho^2$ (dashed line) and up to $g^4/m_\rho^4$ (dotted
line).

The summation (r) of Fig. \ref{fig:resum_schemes} with the
interaction from Eq. (\ref{eff_strong}) exhibits a symmetry factor of $1/N$
for a ring with $N$ ''small'' loops (see Fig. \ref{fig:resum_schemes}) which
after summing over $N$ leads to the occurrence of a logarithmic cut in the zero-component $p^0$ of the
momentum of the ''big'' loop. Due to this obstacle for the contour integration
method \cite{Ka1989}, usually only the static mode $p^0=0$ is calculated,
although new studies overcome this problem \cite{Liao:2002kp}. In the present
approach, we can calculate the ring with $N$ ''small'' loops explicitly before
summing over $N$. This avoids, thus, the problem of the logarithm at the cost of having
to cut the series at some $N_{{\rm max}}$. 
On the positive side, all modes are
included, and not only the $p^0=0$ static contribution.  
The result up to eight ''small'' loops has already converged up
to $T\sim 200$ MeV and is displayed in Fig. \ref{fig:resummed} as (r). The
explicit solution can be found in Appendix \ref{appendixd}.

Note that in the resummation schemes we do not consider the vacuum parts of the loops, i.e. we do not renormalize the vacuum amplitude. 
This excludes potential double counting issues in the final numerical results in Sec. \ref{sec:numres} where resummations and density expansion are added: Renormalizations of the vacuum amplitude are supposed to be be included in the phase shifts that are used in the density expansion.

There is an additional resummation scheme that sums up the $\rho\rho\pi\pi$ interaction required by the gauge invariance of the $\rho$ (see Eq. (\ref{4vertex})):
One can consider diagram (b) and (c) of Fig. \ref{fig:entropies},
dress the $\rho$ propagator as indicated in (t) of Fig. \ref{fig:resum_schemes}, and 
finally take the heavy $\rho$ limit as in Sec. \ref{sec:eff_interaction}.
This leads to the same result as a renormalization of the static $\rho$
propagator $-1/m_\rho^2$ of diagram (a) in Fig. \ref{fig:entropies}
for the $\pi\pi$ interaction in the heavy $\rho$ limit:
The resummed pion tadpoles can be incorporated by a
mass shift,
\be m_{\rho^\pm}^2\to
m_\rho^2+\frac{g^2}{4}(U_++U_-+2D), \quad m_{\rho^0}^2\to
m_\rho^2+\frac{g^2}{2}(U_++U_-)
\label{mrho_effective}
\ee 
for charged and neutral $\rho$. The 
contribution to $m_{\rm el}$ from this modification is shown in 
the insert of Fig. \ref{fig:resummed} as (t). 
The thermal $\rho_0$ mass from
Eq. (\ref{mrho_effective}) at $\mu=0$ is $m_{\rho_0}=824$ MeV at $T=170$ MeV which
is slightly more than in other studies \cite{Sarkar:1997aa}. This is certainly
due to the omission of the $\rho\to \pi\pi\to\rho$ selfenergy which also
contributes and is required by the gauge invariance of the $\rho$-meson. In the 
counting of the present study, the $\rho\to \pi\pi\to\rho$ selfenergy is statically included in the resummation
(n) of Fig. \ref{fig:resum_schemes}.

To the right in Fig. \ref{fig:resummed} the normalized CF over entropy, $D_S$ from Eq. (\ref{observable}), are plotted. For comparison, the result at $g^2$ from the dynamical $\rho$-exchange (see Fig. \ref{first_dqds}) is shown with the dashed line. We include now the resummation (n) but only with three or more loops, or in other words, at $g^4$ and higher in the interaction in order to avoid double counting with the $g^2$ contribution. We have already seen in Fig. \ref{fig:resummed}, left panel, that both resummations (n) and (r) contain the same diagram at order $g^4$ (linear chain of three loops). Thus, again in order to avoid double counting, we include the resummation (r) requiring at least three of the ''small'' loops, see Fig. \ref{fig:resum_schemes}; this means that only contributions of order $g^6$ and higher are included. Finally, we add the resummation (t) including the orders $g^4$ and higher, which again avoids double counting of the $g^2$-contribution. Summing in this way the resummations to the $g^2$-result (dashed line) for both $\ave{\delta Q^2}$ and $S$, the resulting $D_S=\ave{\delta Q^2}/S$ is indicated as $\Sigma'$ with the dashed-dotted line in Fig. \ref{fig:resummed}.

The resummations have a large effect on $\ave{\delta Q^2}$ (see Fig.
\ref{fig:resummed}, left) whereas their effect on the entropy is much smaller;
the entropy is efficiently suppressed for higher orders in the coupling. This
explains, why the result $\Sigma'$ shows such a large difference compared to the results at order $g^2$ (dashed line).

For the resummations (n) and (r), we have ensured that we recover the  results from
Eqs. (\ref{pinullsum}) and (\ref{logmu_eff}) at the same order of the
interaction. We have also verified that the results from Ref. \cite{Gale:1990pn}
at external momentum $p$ of the $\rho$ being zero $(p^0=0, {\bf p}\to 0)$ 
match the $\rho$ self energies at $\mu=0$ that are 
implicitly or explicitly contained in the resummations (n) and (t).

A possible extension of the diagrams discussed here is given by resummations of super-daisy type: 
the pion propagator is dressed by a series of pion tadpoles; the propagator of the tadpole loop itself is again dressed which constitutes a self consistency condition. E.g., this leads to a thermal mass of the pion $m_\pi\sim 170$ MeV at $T\sim 170$
Mev. However, one should realize that the lower orders in the coupling $g$ of a super-daisy expansion are already covered by the resummations considered before: it is easy to see that the super-daisy resummation introduces additional diagrams only at order $g^8$ and higher ($g^6$ and higher for resummation (t)) and, thus, can be neglected.

\subsection{Extension to $\boldsymbol{SU(3)}$}
\label{sec:su3extension}
In order to obtain a more realistic model for the grand canonical partition function,
the leading contributions from the interaction of the full $SU(3)$ meson and vector meson octets is considered.
Obviously, the leading contribution to the CF from strange degrees of freedom 
is simply the free kaon gas. Here we want to discuss corrections due to 
interactions of kaons with pions. The most important of those is the resonant 
$p$-wave interaction involving an intermediate $K^*(892)$ meson. This is quite 
analogous to the $\rho$ meson in the $\pi\pi$ case, discussed previously. The 
$\Phi$ meson, on the other hand, only enters if interactions between kaons 
are considered. These are sub-leading as pions are more abundant and thus 
$\pi K$ interactions are more important. 

As in the previous sections we describe the meson-meson interaction by 
dynamical vector meson exchange, second, by an effective interaction, and, third, by realistic phase shifts via a relativistic Bose-Einstein density
expansion. For processes which contain
at least one pion, the
dynamical vector meson exchange is mediated by the $K^*(892)$. The
effective contact interaction is taken from the LO chiral meson-meson
Lagrangian in its $SU(3)$ version, ${\cal L}_{\pi K}^{(2)}$ for the $\pi K$-interaction. The density expansion of $\pi K$
scattering is obtained following the same steps as in Sec. \ref{sec:relavir}.
Details of the calculations are summarized in Appendix \ref{appendixe}.

\begin{figure}
\includegraphics[width=11.cm]{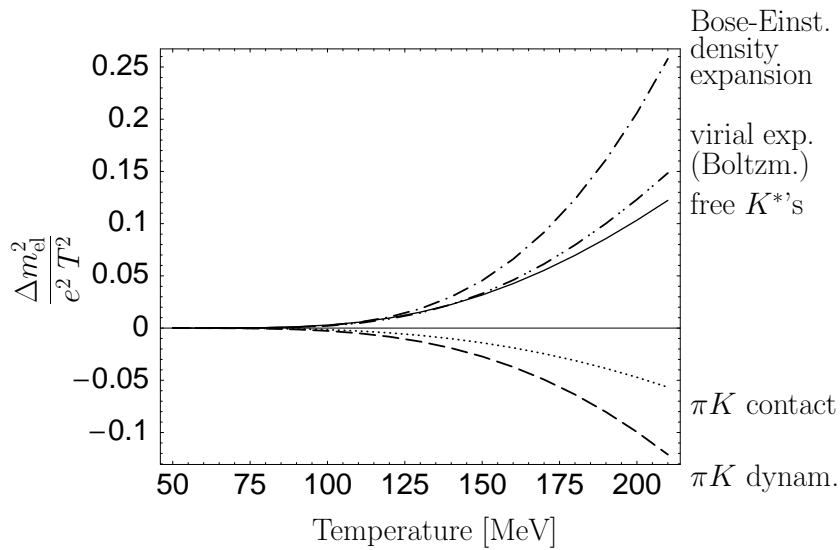} 
\caption{Corrections to the electric mass or 
CF, $\ave{\delta Q^2}/(e^2 V T^3)$ for $\pi K$ interaction. The density and virial expansions are from Eqs. (\ref{pikvirfull}) and (\ref{Boltzmannpik}), respectively. The loop expansions ''$\pi K$ dynamical'' and ''$\pi K$ contact'' are from Eqs. (\ref{logzkstarb},\ref{cdkstar}) and (\ref{logmu_eff_K}), respectively. The solid line shows the electric mass of a gas of free $\kappa(800)$, $K^*(892)$, $K_0^*(1430)$, and
$K_2^*(1430)$ mesons.}
\label{fig:compare_more}
\end{figure}
In Fig. \ref{fig:compare_more} the CF from the different models are shown. For the virial and density expansion the phase shifts have been taken from the parametrization of Ref. \cite{Aston:1987ir} for the attractive channels $\delta_0^{1/2},\delta_1^{1/2},\delta_2^{1/2}$, corrected for the parameters of the $K_0^*(1350)$ resonance (nowadays, $K_0^*(1430)$ in the PDG \cite{Eidelman:2004wy}) as reported in Ref. \cite{Venugopalan:1992hy}. The repulsive $\delta_0^{3/2}$ phase shift is from Ref. \cite{Estabrooks:1977xe}. The phase shifts plotted in Fig. 4 of \cite{Venugopalan:1992hy} up to $s^{1/2}=1$ GeV have been reproduced.

The situation resembles the case of $\pi\pi$ scattering from Fig. \ref{fig:compare_first_kapusta}: Thermal loops with dynamical vector exchange or with effective interaction via ${\cal L}_{\pi K}^{(2)}$ show large discrepancies to the virial and density expansions, this time even more than in the $\pi\pi$ case. The reasons are similar as those found in Sec. \ref{why}: The repulsive $(I,\ell)=(3/2,0)$ partial wave is not well described by $\pi K$ scattering via $K^*(892)$ and unitarity problems of the thermal loops show up. 
The contributions from both the virial expansion and the low density expansion
are large compared to the virial corrections in the $\pi\pi$ sector (see
Fig. \ref{fig:compare_first_kapusta}). This seems surprising as in the $\pi K$
system the kaon has a large mass which should suppress contributions
kinematically. However, in the considered channels of $\pi K$ scattering, four
resonances are present, $\kappa(800)$, $K^*(892)$, $K_0^*(1430)$, and
$K_2^*(1430)$ and we know from Sec. \ref{why} that resonances give a large
positive contribution to $m_{\rm el}$\footnote{Of course in the $\pi\pi$-case
resonances above the $\rho$-meson also contribute. While we have ignored these
in the previous discussion, they will be included in the final analysis given
in the following chapter.}. The electric masses from these resonances, treated as
free gases (Boltzmann), is plotted in Fig. \ref{fig:compare_more} with the solid
line. We find the same pattern as in the discussion of Fig. \ref{fig:tests} for
the free $\rho$: the virial corrections from resonant phase shifts are well
described by a free gas of the same resonances. Furthermore, the repulsive
$\delta_0^{3/2}$ phase shift is very small.

As the outcome for the density expansion in Fig. \ref{fig:compare_more} shows,
the inclusion of Bose-Einstein statistics is important (compare to the virial
expansion which uses Boltzmann statistics). We
consider the density
expansion to provide the most reliable prediction.

\section{Numerical results}
\label{sec:numres}
In the discussions in Secs. \ref{why} and \ref{sec:su3extension} good reasons have been found that at quadratic order in density $n$ the Bose-Einstein density expansion gives the most realistic results. For the final numerical results we include therefore the $\pi\pi$ and the $\pi K$ density expansion from Eqs. (\ref{b2mu}) and (\ref{pikvirfull}). 
\begin{figure}
\includegraphics[width=16.4cm]{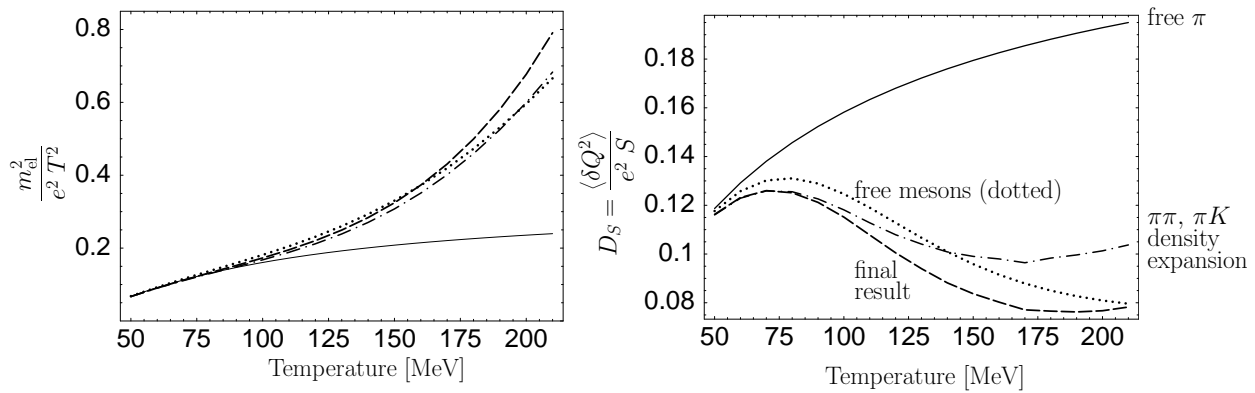}
\caption{Final results for charge fluctuations (electric mass) and $D_S$. Results of the Bose-Einstein density expansions with the dashed-dotted lines. Adding $\rho\rho\pi\pi$ and $K^*K^*\pi\pi$ contributions, the resummations, and free mesons up to 1.6 GeV, the results are indicated with the dashed lines. For comparison, $m_{\rm el}^2$ and $D_S$ from free mesons alone, 
without any interactions, are also plotted (dotted lines).}
\label{fig:add_all}
\end{figure}
The dashed-dotted lines in Fig. \ref{fig:add_all} show electric mass and normalized charge fluctuations $D_S$ from Eq. (\ref{observable}) for the sum of the two density expansions.
At order $n^2$, there are additional photon selfenergy diagrams with
$(\gamma)\gamma\rho\rho$ and $\rho\rho\pi\pi$ vertices from Fig.
\ref{fig:overview_more}. As discussed at the end of Sec. \ref{why} these
contributions are not included in the density expansion but a consequence of the
$\rho$ being introduced as a heavy gauge field. The same applies to the
$K^*K^*\pi\pi$ diagram (d) from Eq.
(\ref{cdkstar}). Thus, we include these additional contributions for
$\ave{\delta Q^2}$ and $S$.

At higher orders in density one has to rely on resummation schemes. 
Including the resummations in the final results does not to lead to double
counting: Resummations at order $g^4$ and
upwards in the strong coupling correspond to diagrams with three and more
loops and, thus, to contributions higher than quadratic in density. We include
(with $g^4$ and higher) the summations (n), (r), and (t) from Sec.
\ref{sec:necklace}. Note that for $m_{\rm el}$ there is a partial cancellation of sizable contributions from the resummations and the $(\gamma)\gamma\rho\rho$, $\rho\rho\pi\pi$, $K^*K^*\pi\pi$ diagrams.

In order to obtain a more realistic picture, we also include as free gases all mesons from the PDG \cite{Eidelman:2004wy} which have not been considered so far, up to a mass of 1.6 GeV. Note that we do not add free mesons that have the same quantum numbers as the density expansions namely $\sigma(600)$, $\rho(770)$, $\kappa(800)$, $K^*(892)$, $K_0^*(1430)$, and
$K_2^*(1430)$. We have seen in Sec. \ref{why} that their contribution via phase shifts in the density expansions is roughly of the size as if they had been included as free particles. Adding all contributions mentioned, the results are indicated with the dashed lines in Fig. \ref{fig:add_all}.

Compared to the density expansions the final results do not change much. The
influence of heavier mesons than those considered
in this study is, thus, well controlled. Many of the heavier resonances that
have been added here as free gases are axials which decay into three particles.
To include them in a density expansion would require the consistent treatment
of three body correlations. 

Concluding, we can assign $D_S \simeq 0.09$ for temperatures $120<T<200$ MeV
which coincides (incidentally) quite well with the result
if one simply considers free, noninteracting, mesons up to masses of $1.6$ GeV.
The latter case is indicated with the dotted lines in Fig. \ref{fig:add_all}.

Theoretical uncertainties in the present study arise from the omission of diagrams such as the (small) eye shaped diagram mentioned in Ref. \cite{Eletsky:1993hv} already at $g^4$. Furthermore,
both resummations and density expansions are incomplete as they only partly include the in-medium renormalization of the resonances which drive the meson-meson scattering, such as the $\sigma(600)$, $f_0(980)$, or the $\rho(770)$ itself \cite{Dashen:1974jw,Gale:1990pn}. In this context one can think of a more complete microscopical model: We have found in Sec. \ref{sec:rho_dynamical} and \ref{why} that unitarity and a good description of the vacuum data up to high energies and in all partial waves are important. Such models exist, e.g., the chiral unitary approach from Ref. \cite{Oller:2000ma}. The medium implementation of such a model has been done in a different context, see \cite{Dobado:2002xf,GomezNicola:2002an} and references therein. 
A generalization of the virial expansion from Ref. \cite{Pelaez:2002xf}
to finite chemical potential and including Bose-Einstein statistics, as carried
out here, would be feasible in principle. Such an ansatz \cite{in_preparation}
would allow to take simultaneously into account the medium
renormalization of the (dynamically generated) resonances and the calculation of
the grand canonical partition function at finite $\mu$ as needed for a
calculation of $m_{\rm el}^2$.

\section{Summary and Conclusions}
For an estimate of charge fluctuations (CF) in the hadronic phase of heavy ion collisions, we have calculated the effect of particle
interactions. For the perturbative expansion up to two thermal loops, the $\pi\pi$ interaction has been described by
vector meson exchange. The correlations induced by a dynamical $\rho$ have
been found significant by comparing to an effective theory where the $\rho$ is frozen out. 

The photon self energies are charge conserving and shown to be equivalent to the loop expansion
of the grand canonical partition function at finite chemical potential. We have
pointed out that the inclusion of imaginary parts is essential for 
a proper description of the thermodynamics, especially if
resonant amplitudes are involved. To second order in the density, it has been possible to include
Bose-Einstein statistics in the conventional virial expansion. This ''density expansion'' can change
the conventional results significantly. Moreover, for real amplitudes, we could
show the equivalence of the loop expansion and 
the density expansion at all temperatures. However, the inclusion of unitary
(complex) amplitudes is more straightforward in the density (virial) expansion.
To the extend that
two-particle correlations are dominant, the density expansion with Bose-Einstein
statistics is, thus, the method of choice as it provides the same statistics as
the thermal loop expansion and unitarity is automatically implemented by the use
of realistic phase shifts.

For an estimate of three- and higher particle correlations, a variety of
summation schemes has been presented, all of which tend to soften the large
first order correction of the thermal loop expansion. 
For the CF, higher order corrections have a large 
influence whereas higher orders for the entropy are small. 

For the CF over entropy, $D_S$, it has been shown that the influence of heavy
particles beyond the interactions considered are well under control; a final
value of $D_S \simeq 0.09$ has been found for temperatures
$120<T<200$ MeV. This result agrees quite well with the outcome
from the free resonance gas, supporting the notion that resonant amplitudes
dominate the thermodynamics. As lattice gauge calculations with realistic quark
masses become available it would be interesting to see at which point these start 
to significantly deviate from a hadron gas.

\ \\
\noindent {\bf Acknowledgments:} This work was supported by the Director, 
Office of Science, Office of High Energy and Nuclear Physics, 
Division of Nuclear Physics, and by the Office of Basic Energy
Sciences, Division of Nuclear Sciences, of the U.S. Department of Energy 
under Contract No. DE-AC03-76SF00098. It has also been supported by the {\it Studien\-stiftung des Deutschen Volkes} and the program {\it Formaci\'on de Profesorado Universitario} of the Spanish Government.

\appendix
\section{From charge fluctuations to photon selfenergy in sQED}
\label{appendixa}
In this section, an outline for the proof of Eq. (\ref{connection}) for scalar
QED is given. The argument follows Ref. \cite{Ka1989} where a
similar connection is made for QED. If $\pi\pi$ contact interactions are
included according to Eq. (\ref{effL}), the steps outlined below are similar,
but lengthier, and Ward identities for four-point functions have to be
determined.

CF are defined as $\ave{\delta Q^2}=\ave{Q^2}-\ave{Q}^2$, and
the expectation values are calculated via the statistical operator of the
grand canonical ensemble with the charge chemical potential $\mu\equiv\mu_Q$. One
obtains immediately: 
\be \langle\delta Q^2\rangle=e^2 TV\;
\frac{\partial}{\partial \mu} \langle\hat{{\mbox {\j}}}_0\rangle \ee 
with
$\hat{{\mbox {\j}}}_0$ the zero-component of the conserved current,
$\hat{Q}=\int \hat{{\mbox {\j}}}_0=V\hat{{\mbox {\j}}}_0$. The expectation
value of
$\hat{{\mbox {\j}}}_0=i(\phi^\star(\partial^0+ieA^0)\phi-\phi(\partial^0-ieA^0)\phi^\star)$ can
be expressed in terms of the propagator
\be \left(\frac{\partial \langle
\hat{{\mbox {\j}}}_0\rangle}{\partial \mu}\right)_T =
-\frac{\partial}{\partial \mu}T
\sum_{\omega_n=-\infty}^{\infty}\int\frac{d^3{\bf p}}{(2\pi)^3}
\;2(p^0-\mu)\;{\cal G}(p^0,{\bf p}) \ee 
where we have used $\mu=eA^0$ and
the definition of the imaginary time propagator 
\be {\cal
G}_{\alpha\beta}({\bf x}\tau;{\bf x}'\tau')=-{\rm Tr}[{{\hat \rho}_G \;{\rm
T}_\tau[\phi_{K\alpha}({\bf x}\tau)\phi^\dagger_{K\beta}({\bf x}'\tau')]]} 
\ee
where $T_\tau$ is the $\tau$-ordered product in the modified Heisenberg
picture, see, e.g., Ref. \cite{FeWa}, and the Fourier transform is at equal
times $\tau$, $\tau^+$ and position ${\bf x=x'}$. The $\mu$-dependence of the propagator is given by
$p^0=i\,\omega_n-\mu$ where $\omega_n=2\pi i\,n\,T$. With this, the derivative can be rewritten as 
\be
\left(\frac{\partial \langle \hat{{\mbox {\j}}}_0\rangle}{\partial
\mu}\right)_T=-\sum_{\omega_n=-\infty}^{\infty}T\int\frac{d^3{\bf
p}}{(2\pi)^3} \left(-2{\cal G}(p^0,{\bf
p})-2p^0\;\frac{\partial}{\partial p_0}\;{\cal G}(p^0,{\bf
p})\right)
\label{appa:secondstep}
\ee 
at zero chemical potential $\mu=0$. Using $\partial/\partial p^0 {\cal
G}=-{\cal G}(\partial/\partial p^0 {\cal G}^{-1}){\cal G}$, the Ward identity
in the differential form for scalar QED can be applied. The Ward identity connects
the inverse propagator with the fully dressed vertex $\Gamma^\mu$ according to
\begin{eqnarray}
e^2TV\left(\frac{\partial \langle \hat{{\mbox {\j}}}_0\rangle}{\partial
\mu}\right)_T&=&T^2V\sum_{\omega_n=-\infty}^{\infty}
\int\frac{d^3{\bf
p}}{(2\pi)^3} \Big[2e^2{\cal G}(p^0,{\bf p}) -e(2p^0){\cal G}(p^0,{\bf
p})\Gamma^0({\bf p}, \omega_n) {\cal G}(p^0,{\bf p})\Big]
\nonumber \\ \nonumber \\ &=&TV\big(\underbrace{\Pi^{00}_{D\;{\rm
mat}}(k_0=0,{\bf k}\to{\bf 0})}_{\begin{picture}(10,40) \put(-40,0){
\includegraphics[width=3cm]{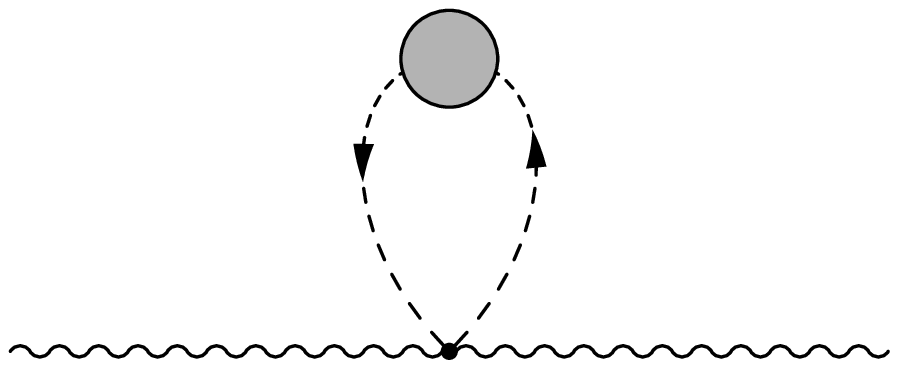}}
\end{picture}
} + \underbrace{ \Pi^{00}_{C\;{\rm mat}}(k_0=0,{\bf k}\to{\bf 0})}_{
\begin{picture}(10,40)
\put(-40,0){ \includegraphics[width=3cm]{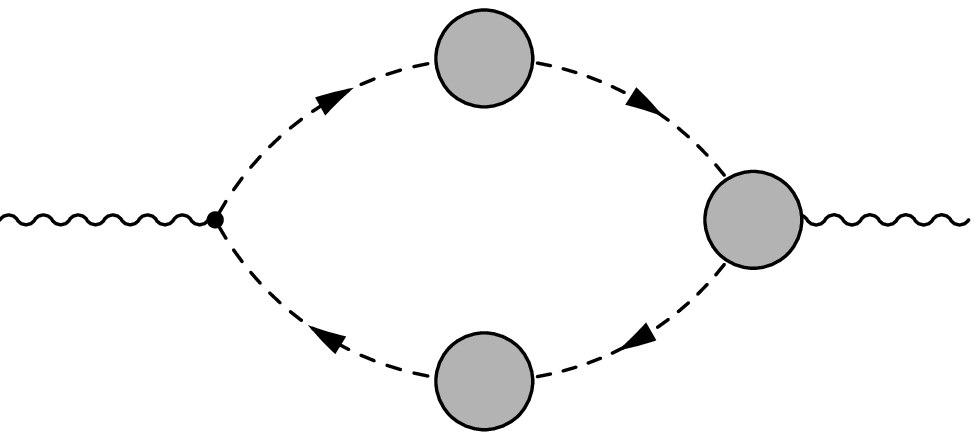}}
\end{picture} 
} \big).
\label{final_sQED}
\end{eqnarray}
Factors of $e$ and $p^0$ have been identified here with the bare
$\gamma\gamma\pi\pi$ and $\gamma\pi\pi$ vertices. In the step from Eq. (\ref{appa:secondstep})
to Eq. (\ref{final_sQED}) we have generated three
propagators from one, and it should be noted that this takes place inside
the momentum integral and summation. Therefore, the limit in Eq.
(\ref{connection}) has
to be taken {\it before} summation and integration.

\section{Pion-pion interaction}
\subsection{Chiral $\boldsymbol{\pi\pi}$ interaction and vector exchange}
\label{appendixb}
In this section the effective pion-pion contact interaction from Sec.  \ref{sec:eff_interaction}
and its connection to the chiral Lagrangian is discussed in more detail. 
For four pion fields the kinetic term of ${\cal L}_{\pi\pi}^{(2)}$ in Eq. (\ref{chiral}) and 
the effective interaction in Eq. (\ref{eff_strong}) have identical isospin and momentum structure.
Comparing the overall coefficients leads to the 
result in Eq. (\ref{ksfr}) which differs from the KSFR relation by a factor of $3/2$.
Studying the low energy behavior of both theories helps solve this puzzle of the obvious violation 
of the phenomenologically well-fulfilled KSFR relation. The $\pi\pi$ amplitude at threshold
from the LO chiral Lagrangian Eq. (\ref{chiral}) and the effective interaction Eq. (\ref{eff_strong}) 
is given by 
\be
T_{\pi\pi}^{(2)}=-\;\frac{2m_\pi^2}{f_\pi^2},\quad T_{{\rm eff}}=-\;\frac{4 g^2 m_\pi^2}{m_\rho^2},
\ee
respectively which leads to the correct KSFR relation
\be
2f_\pi^2 g^2=m_\rho^2.
\label{ksfr_correct}
\ee
This is due to the mass correction
term proportional to ${\cal M}$ in Eq. (\ref{chiral}). This term, however,
does not have any momentum structure and immediately becomes small at finite
pion momenta compared to the kinetic term. It has no influence in the results
of this study.

For finite pion momenta, higher order partial waves have to be included.
We concentrate on the quantum numbers of the $\rho$-meson and obtain for 
$\pi\pi$ scattering via the LO chiral interaction in isospin $I=1$:
\be
T^1_{\pi\pi}=\frac{-1}{f_\pi^2}\;(t-u)
\label{t11chiral}
\ee
which should be compared to the result from $\rho$-exchange from Eq. (\ref{34vertex}):
\be
T^1({\rm dyn.}\;\rho)&=&g^2\left(\frac{s-u}{t-m_\rho^2}+2\;\frac{t-u}{s-m_\rho^2+i m_\rho\Gamma(s)}+\frac{t-s}{u-m_\rho^2}\right),\non
T^2({\rm dyn.}\;\rho)&=&g^2\left(\frac{u-s}{t-m_\rho^2}+\frac{t-s}{u-m_\rho^2}\right)
\label{t1_rho}
\ee 
where we have also given the result for $T^2$ for completeness, and $T^0$
is immediately obtained by crossing symmetry, $T^0=-2T^2$.
Projecting out the $p$-wave in both results (\ref{t11chiral}) and (\ref{t1_rho}) by using
\be
T^I_\ell(s)=\frac{1}{64\pi}\int_{-1}^1d(\cos\theta)\;P_\ell(\cos\theta)\;T^I(s,t,u)
\label{pwaintegral}
\ee
for $(I,\ell)=(1,1)$,
making an expansion in ${\bf p}_{\rm cm}^2$, 
and comparing the coefficients, leads to the relation
$
m_\rho^2-4m_\pi^2=3f_\pi^2g^2
$
which shows again the deviation of 3/2 from the KSFR relation up to a correction from the pion mass. 
However, taking only the $s$-channel vector 
exchange, which is given by the second term
of $T^1$ Eq. (\ref{t1_rho}), we obtain after projection to the $p$-wave:
\be
m_\rho^2-4m_\pi^2=2f_\pi^2g^2.
\ee
This is indeed the KSFR relation in Eq. (\ref{ksfr_correct}) with some small correction which vanishes when 
$s$ is neglected against $m_\rho^2$ in the denominator of Eq. (\ref{t1_rho}).
Concluding, the restriction to $s$-channel vector exchange in $\pi\pi$ scattering restores the
KSFR relation in the $p$-wave expansion of the scattering amplitude. 
However, $t$- and $u$-channel vector exchange is also present, and this
leads to the effective interaction in Eq. (\ref{eff_strong}) which is 3/2 times stronger
than the interaction from the LO chiral Lagrangian.

Fig. \ref{fig:pipivac} illustrates the behavior of 
the different theories together with data from Ref. \cite{Froggatt:1977hu}:
\begin{figure}
\includegraphics[width=7cm]{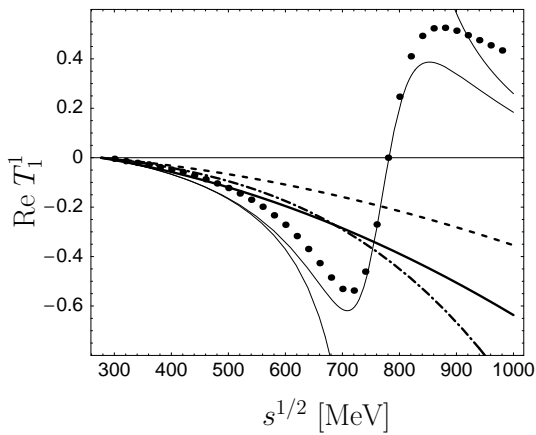}
\caption{$p$-wave isovector $\pi\pi$ interaction. Dots: Partial 
wave analysis from Ref. \cite{Froggatt:1977hu}. 
Dashed line: ${\cal L}_{\pi\pi}^{(2)}$ calculation. 
Dashed-dotted line: ${\cal L}_{\pi\pi}^{(4)}$ calculation from Ref. \cite{Donoghue:1992dd}. 
Solid line: Effective interaction from Eq. (\ref{eff_strong}). 
Thin solid lines: Explicit $\rho$ exchange from Eq. (\ref{34vertex}) 
with and without (momentum dependent) width for the $\rho$.}
\label{fig:pipivac}
\end{figure}
The LO chiral Lagrangian underpredicts the strength of the experimental $T_1^1$ amplitude.
In contrast, the interaction up to ${\cal L}_{\pi\pi}^{(4)}$ 
and the effective interaction from Eq. (\ref{eff_strong})
describe better the data at low energies. The explicit $\rho$
exchange with width (thin line) delivers a good data description even
beyond the $\rho$-mass.

One more remark is appropriate in the framework of this
section: 
In the treatment of the $\rho$-meson as a heavy gauge field, the covariant
derivative 
introduces the $\pi\rho$ interaction as we have seen in Sec. \ref{sec:eff_interaction}.
Additionally, the original $\pi\pi$ interaction from Eq. (\ref{chiral}) remains in 
this derivative. In the present model, we have omitted this term, as has been
also done, e.g., in Ref. \cite{Klingl:1996by}. 
This leads to better agreement with the data in the $T_1^1$-channel and ensures
the KSFR relation.
It is possible to keep the original chiral interaction,
but then additional refinements have to be added added as, e.g., in Ref. \cite{Cabrera:2000dx}.

\subsection{Unitarization of the $\boldsymbol{\pi\pi}$-amplitude with the $\boldsymbol{K}$-matrix}
\label{appendixb1}
The
$K$-matrix is defined via the $S$-matrix as
\be
S_K(E)=\frac{1+i\,Q\,K}{1-i\,Q\,K},\quad K=-\,\frac{T_{\rm tree}}{16\pi\,E}
\ee
with the tree level amplitude $T_{\rm tree}$ from Eq. (\ref{t11chiral}) and the c.m. momentum $Q$ from Eq. (\ref{boostbose}). 
The unitarized amplitude $T_{(u),1}^1$ which is given by
\be
T_{(u),1}^1=\frac{T_1^1}{1+2i\,Q\,T_1^1/E},\quad T_1^1=-\frac{E^2-4m_\pi^2}{96\pi\, f_\pi^2}
\label{tree_level_l2}
\ee
can be parametrized via phase shift as
\be
\delta_1^1=\frac{1}{2}\arctan\;\frac{-{\rm Re}\,T_{(u),1}^1}{\frac{E}{4Q}+{\rm Im}\,T_{(u),1}^1}.
\label{unit}
\ee

\section{The $\boldsymbol{\rho}$-meson in the heatbath}
\label{appendixcc}
\subsection{Analytic results}
\label{appendixc}
The analytical expressions and numerical contributions from the set of gauge
invariant diagrams in Fig. \ref{fig:overview_2loop} are given which are obtained 
from the interactions from Sec. \ref{sec:eff_interaction}. With \be
L_\pm\left(a,b\right):=\log\left|\frac{\left[m_\rho^2+\left(p\pm
q\right)^2-\left(b\;\omega-a
\omega'\right)^2\right]\left[m_\rho^2+\left(p-q\right)^2-\left(b\;\omega+a
\omega'\right)^2\right]}{\left[m_\rho^2+\left(p\mp
q\right)^2-\left(b\;\omega-a
\omega'\right)^2\right]\left[m_\rho^2+\left(p+q\right)^2-\left(b\;\omega+a
\omega'\right)^2\right]}\right|,
\label{def_L}
\ee 
where $\omega^2=q^2+m_\pi^2$ and $\omega'^2=p^2+m_\pi^2$, we obtain for the real parts of the diagrams in
Fig. \ref{fig:overview_2loop}, left column: \be \Pi_{({\rm
1a1})}^{00}(k_0=0,{\bf k\to 0})&=&\;\frac{-e^2g^2}{\left(2\pi\right)^4
m_\rho^2}\;\frac{\partial}{\partial\alpha}\;\frac{\partial}{\partial\beta}|_{\alpha=\beta=1}
\int\limits_0^{\infty}dp\int\limits_0^\infty
dq\;\frac{pq}{\alpha\beta\omega\omega'^3}\;n[\omega]\;n\left[\alpha/\beta\;\omega'\right]\non
&\times&\left[-8m_\rho^2\; p q+\left[(2 m_\pi
m_\rho)^2-\left(m_\rho^2-\left(\left(\alpha/\beta\right)^2-1\right)\omega'^2\right)^2\right]L_-\left(\alpha/\beta,1\right))\right],\non
\Pi_{({\rm 1a2})}^{00}(k_0=0,{\bf k\to
0})&=&\;\frac{-e^2g^2}{2\left(2\pi\right)^4}\;
\;\frac{\partial}{\partial\alpha}\;\frac{\partial}{\partial\beta}|_{\alpha=\beta=1}
\int\limits_0^{\infty}dp\int\limits_0^\infty
dq\;\frac{pq}{\left(\omega\omega'\right)^2}\;n[\beta \omega]\;n\left[\alpha
\omega'\right]\non
&\times&\left[4m_\pi^2-m_\rho^2+2\left((\beta^2-1)\omega^2+(\alpha^2-1)\omega'^2\right)-\frac{1}{m_\rho^2}\left(
(\alpha^2-1)\omega'^2-(\beta^2-1)\omega^2\right)^2\right]\non
&\times&L_+\left(\alpha,\beta\right),\non \Pi_{({\rm 4a})}^{00}(k_0=0,{\bf k\to
0})&=&\;\frac{-e^2g^2}{\left(2\pi\right)^4
m_\rho^2}\;\frac{\partial}{\partial\alpha}|_{\alpha=1}\int\limits_0^{\infty}dp\int\limits_0^\infty
dq\;\frac{pq}{\alpha\omega\omega'^3}\;n[\omega]\;n[\alpha\omega']\non
&\times&\left[-8m_\rho^2\; p q+\left[(2 m_\pi
m_\rho)^2-\left(m_\rho^2-(\alpha^2-1)\omega'^2\right)^2\right]L_-(\alpha,1)\right],\non
\Pi_{({\rm 5a})}^{00}(k_0=0,{\bf k\to 0})&=&\frac{6
e^2g^2}{\left(2\pi\right)^4 m_\rho^2}
\;\frac{\partial}{\partial\alpha}|_{\alpha=1}\;
\int\limits_0^{\infty}dp\int\limits_0^\infty
dq\;\frac{pq}{\omega\omega'^2}\;n[\omega]\;n\left[\alpha \omega'\right]\non
&\times&\left[-\alpha\omega'\left(m_\rho^2+(1-\alpha^2)\omega'^2\right)L_-\left(\alpha,1\right)
+\omega\left(m_\rho^2+(\alpha^2-1)\omega'^2\right)L_+\left(\alpha,1\right)
\right],\non \Pi_{({\rm 6a})}^{00}(k_0=0,{\bf k\to 0})&=&\;\frac{-3
e^2g^2}{\left(2\pi\right)^4 m_\rho^2}\;
\int\limits_0^{\infty}dp\int\limits_0^\infty
dq\;\frac{pq}{\omega\omega'}\;n[\omega]\;n\left[\omega'\right]\non
&\times&\left[-\left(m_\rho^2-\omega^2-\omega'^2\right)L_-\left(1,1\right)
+2\omega\omega'L_+\left(1,1\right) \right],\non \log Z_{{\rm
(en1)}}&=&\frac{3g^2 V}{4T\left(2\pi\right)^4}\;
\int\limits_0^{\infty}dp\int\limits_0^\infty
dq\;\frac{pq}{\omega\omega'}\;n[\omega]\;n\left[\omega'\right]\left[-8 p
q+\left(4 m_\pi^2-m_\rho^2\right)L_-\left(1,1\right))\right].
\label{analytic_2_loop}
\ee 
In these expressions the poles from the $\rho$-propagator have been
omitted as discussed in Appendix \ref{appendixc1}.  The use of derivatives
in Eq. (\ref{analytic_2_loop}) cures infrared divergences which occur (see Appendix \ref{appendixc1}).
The logarithmic pole in the
numerically relevant integration regions for $p$ and $q$ is in all cases is
given by 
\be
m_\rho^2+(p-q)^2-\left(b\; \omega+a\;\omega'\right)^2=0
\label{log_pole_position}
\ee 
where $a,b$ take values according to the arguments of
$L_\pm(a,b)$ of Eqs. (\ref{def_L},\ref{analytic_2_loop}). The
singularity leads to an imaginary part which we neglect. 
The issue of imaginary parts is discussed in Sec. \ref{why}.
The diagrams from the second column of Fig. \ref{fig:overview_2loop} are calculated straightforward
with the results
\be
\Pi^{00}_{(1c1)}=-\frac{e^2g^2}{m_\rho^2}\;C^2, \quad \Pi^{00}_{(5c1)}=-\frac{2e^2g^2}{m_\rho^2}\;CD, \quad
\Pi^{00}_{(6c)}=-\frac{e^2g^2}{m_\rho^2}\;D^2
\label{easyloops}
\ee
in the static limit ($k_0=0,{\bf k\to 0}$).

\begin{figure}
\includegraphics[width=16cm]{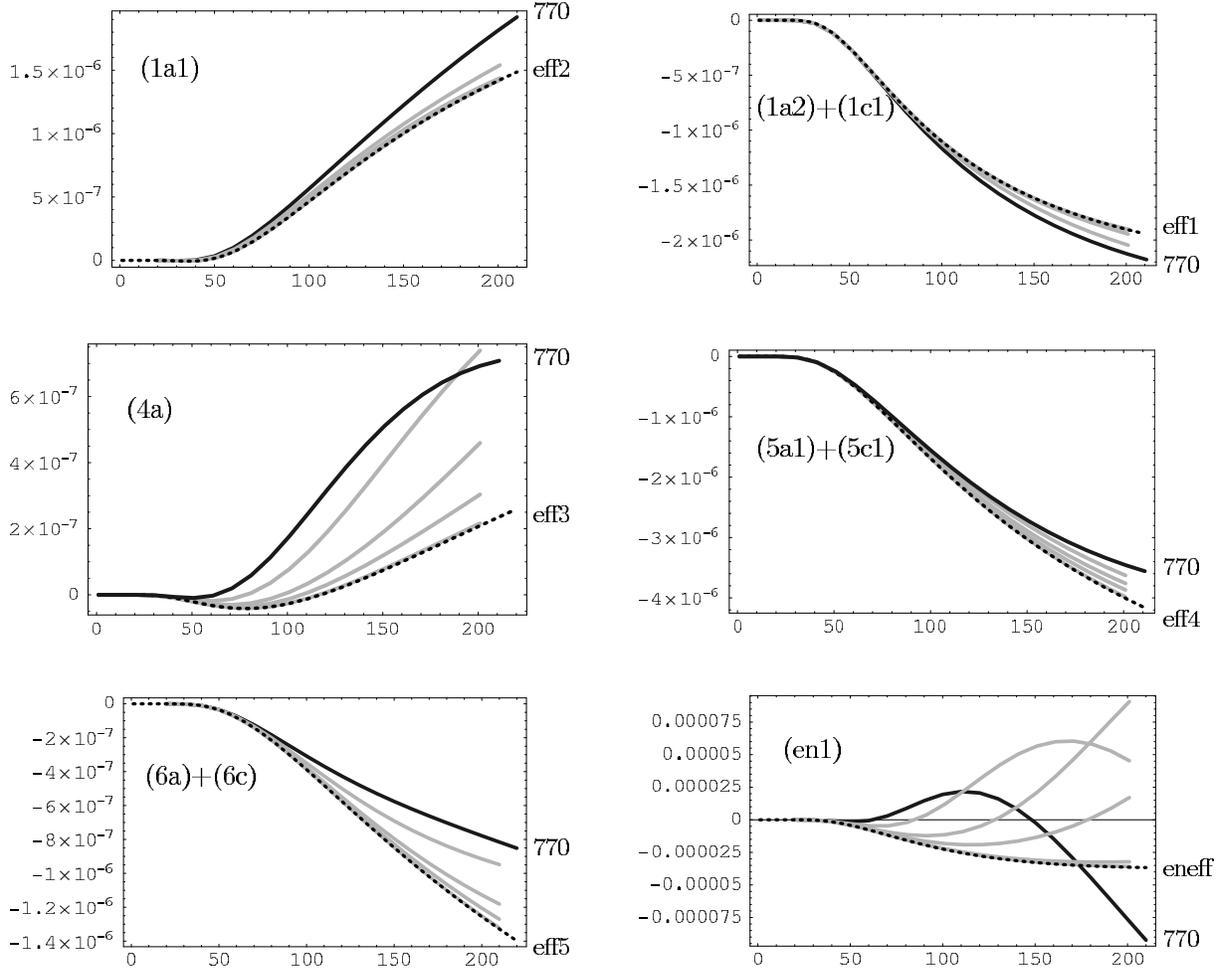} 
\caption{Numerical results for the diagrams from Fig. \ref{fig:overview_2loop}, 
Eq. (\ref{analytic_2_loop}) as a function of $T$ [MeV]. Selfenergy $\Pi/(e^2T^4)$ 
in [MeV$^{-2}$] for all plots, except the correction to $Z$: 
$\log Z_{{\rm (en1)}}/(V T^4)$ in [MeV]$^{-1}$. Results for different 
$m_\rho$ with solid lines. Dashed lines: Corresponding diagrams from the 
heavy $\rho$ limit, see Tab. \ref{tab:num_res_eff} 
and Eq. (\ref{s_1}).}
\label{fig:num2l}
\end{figure}

Fig. \ref{fig:num2l} shows the numerical results.
For every diagram, the contribution of the dynamical $\rho$-meson at
its physical mass of $m_\rho=770$ MeV (indicated with ''770'') is displayed.
 Additionally,
the amplitudes for $\rho$-masses of $m_\rho^i=1070,1770, 2770, 10000$ MeV are
evaluated, multiplying the result with $(m_\rho^i/m_\rho)^2$ (gray
lines). This would correspond to a $\rho$-meson with mass $m_\rho^i$ whose
strong coupling $g$ is increased by $(m_\rho^i/m_\rho)$. This is indeed
equivalent to the heavy $\rho$ limit from Sec.
\ref{sec:eff_interaction} and convergence of the results from
Eq. (\ref{analytic_2_loop}) towards the heavy $\rho$ limit of Sec. \ref{sec:effective_results} 
(dashed lines) is
observed. This convergence is, on the other hand, a useful tool to check the
results from Eq. (\ref{analytic_2_loop}).

The large difference of both models at $m_\rho=770$ MeV in case of the diagrams (4a) and (eff3)
is due to terms that partially cancel: diagram
(eff3) $\sim D (2D-C)$. For the calculation
of the entropy in Eq. (\ref{observable}), the correction to $\log Z$ is needed  
which is very different
for diagram (eneff) and diagram (en1) as Fig. \ref{fig:num2l} shows. 
The discrepancy can be traced back to the different high energy behavior of the amplitudes.
In any case, the total size of the entropy correction, compared to the result 
of the free pion gas, Eq. (\ref{free_entropy}), is small and of no relevance
for the final results.

\subsection{Calculation of diagram (1a2)}
\label{appendixc1}
The calculation of one of the diagrams from Fig. \ref{fig:overview_2loop} is
outlined in more detail. The evaluation of the other diagrams is carried
out in an analog way, with the results given in Eqs. (\ref{analytic_2_loop},\ref{easyloops}).
For diagram (1a2) it is most convenient to treat the vertex correction first,
that is given by the left side of the diagram. The external photon momentum has to be set to
zero from the beginning of the calculation as has been shown in Appendix \ref{appendixa};
the matter part of the vertex correction
reads for an external $\pi^+$: 
\be \Gamma^0[k^0=0,{\bf k \to 0}]&=&\frac{2e
g^2 p^0}{\pi^2}\;\int\limits_0^\infty dq\;q^2\int\limits_{-1}^{1} dx
\;\frac{1}{2\pi i}\int\limits_{-i\infty+\epsilon}^{i
\infty+\epsilon}dq^0\;n[q^0]\;\frac{\left(q^0\right)^2}{\left((q^0)^2-\omega^2\right)^2}\non
&\times&\frac{4m_\pi^2-m_\rho^2+2\left((p^0)^2+(q^0)^2-\omega^2-\omega'^2\right)-
\left(1/m_\rho^2\right)\left((p^0)^2-(q^0)^2+\omega^2-\omega'^2\right)^2}{\left((p^0+q^0)^2-\eta^2\right)\left((p^0-q^0)^2-\eta^2\right)}
\label{vertex_correction}
\ee 
where the contour integration method of Ref. \cite{Ka1989} is used for the
summation over Matsubara frequencies. In Eq. (\ref{vertex_correction}),
$\omega^2=q^2+m_\pi^2$ and $\eta^2=p^2+q^2-2pqx+m_\rho^2$. 
A problem occurs when closing the
integration contour in the right $q^0$ half plane: The residue at $\omega$
from the double pole of the two pion propagators at the same energy 
is given by 
\be {\rm Res}\; f(z)|_{z=\omega}=\lim_{z\to
\omega}\frac{1}{(m-1)!}\frac{d^{m-1}}{dz^{m-1}}[f(z)(z-\omega)^m]
\label{resformula}
\ee at $m=2$. The derivative also applies to the denominator in the second
line of Eq. (\ref{vertex_correction}) from the $\rho$ propagator. The
integrand exhibits then a divergence of the type 
\be \Gamma[k^0=0,{\bf k \to 0}]\sim\int dq\;\frac{1}{a-q^2}, \quad\mbox{at}\quad p^0=0.  \ee 
The divergence
affects only the zero-mode $p^0=0$, but when the pion lines are closed later
on, in order to obtain diagram (1a2), the integrals in
Eq. (\ref{vertex_correction}) are not defined any more, and one finds poles of
the type $1/(a-{\bf q}^2)$ in the three-momentum integration.  This infrared
divergence, for the external photon at $k=0$, occurs in diagrams that contain,
besides two or more propagators at the same momentum, an additional propagator
as in this case the one of the $\rho$-meson. 

The complication can be most easily
overcome with the introduction of additional parameters according to 
\be
\frac{1}{2\omega^2}\frac{\partial}{\partial\beta}|_{\beta=1}\;\frac{1}{(q^0)^2-
(\beta\omega)^2}=\frac{1}{\left((q^0)^2-\omega^2\right)^2},
\label{trickder}
\ee 
and performing the derivative numerically after the three-momentum
integration. Still, singularities of the $1/q$ type remain, but they are
well-defined by the $\epsilon$-prescription in the $q^0$-integral of
Eq. (\ref{vertex_correction}). We can in this case, as well as in all other
diagrams from Fig. \ref{fig:overview_2loop}, integrate the angle $x=\cos({\bf
p,q})$ analytically, thus being left with logarithmic singularities, that are
easily treated numerically with the help of Eq. (\ref{log_pole_position}).

It has been checked for all diagrams in Fig. \ref{fig:overview_2loop} that the
poles of the $\rho$-meson can be omitted: In 
Eq. (\ref{vertex_correction}) the denominator of the second line from the
$\rho$-propagator produces two single poles in the right $q^0$ half
plane. Taking these residues into account in the contour integration leads to
deviations of less than 1 \% of the result for the vertex correction, for all
values of $(p^0,{\bf p})$ and up to temperatures $T\sim 200$ MeV. Intuitively,
this is clear since these poles produce a strong Bose-Einstein suppression
$\sim n[m_\rho]$ and extra powers of $m_\rho$ in the denominator compared to
the pion pole. This approximation is made for all results of
Eq. (\ref{analytic_2_loop}). See also Sec. \ref{why} where the approximation is again tested.

The rest of the evaluation of diagram (1a2) is straightforward up to the
introduction of an additional derivative parameter in the same manner as
above. As one can see in Fig. \ref{fig:overview_2loop}, a topologically
different structure, diagram (1c1), is possible for the combination of two
$\gamma\pi\pi$ and two $\rho\pi\pi$-vertices. This diagram is easily
evaluated and has to be added.

\subsection{The $\boldsymbol{\gamma\pi\rho}$ system at finite $\boldsymbol{\mu}$.}
\label{appendixc2}
Explicit results for $\log Z$ from the diagrams (b), (c), and (d) from Fig. \ref{fig:entropies} are given
from which the electric mass can be directly calculated using Eq. (\ref{stat_mech_cf}). As argued in the main text, the diagrams (b,c,d) from Fig. \ref{fig:entropies} lead to the same CF as all diagrams with dynamical $\rho$ from Figs. \ref{fig:overview_2loop} and \ref{fig:overview_more}.
For diagram (b), the result is
\be \log Z^{\pi\pi}_{{\rm
(b)}}(\mu)&=&\frac{-g^2\beta V}{32}(U_++U_-)(U_++U_-+4D) +\frac{g^2\beta
V}{128\pi^4}\int\limits_0^\infty dp\int\limits_0^\infty
dq\;\frac{pq\left(4m_\pi^2-m_\rho^2\right)}{\omega\omega'}\non
&\times&\big[\left(n_+n[\omega'-\mu]+n_-n[\omega'+\mu]\right)\log_1+
\left(n_+n[\omega'+\mu]+n_-n[\omega'-\mu]\right)\log_2\big]
\label{logb}
\ee 
with $U_\pm$ and $V_\pm$ from Eq. (\ref{u_v}), 
$n_\pm=n[\omega\pm\mu]+2n[\omega]$, and
\be
\log_1=\log\left[\frac{m_\rho^2+(p-q)^2-(\omega+\omega')^2}
{m_\rho^2+(p+q)^2-(\omega+\omega')^2}\right],\;
\log_2=\log\left[\frac{m_\rho^2+(p-q)^2-(\omega-\omega')^2}
{m_\rho^2+(p+q)^2-(\omega-\omega')^2}\right],
\label{log1log2}
\ee 
with $\omega^2=q^2+m_\pi^2$, $\omega'^2=p^2+m_\pi^2$, 
and $n$ the Bose-Einstein distribution. We have checked that $\log Z_{{\rm
(b)}}(\mu=0)=\log Z_{{\rm (en1)}}$ from Eq. (\ref{analytic_2_loop}). The
diagram (c) in Fig. \ref{fig:entropies} with 
\be \log Z^{\pi\pi}_{{\rm
(c)}}(\mu)=-\;\frac{g^2\beta V}{8m_\rho^2}\left(V_+-V_-\right)^2
\label{logc}
\ee 
is zero for $\mu=0$ and therefore $\log Z_{{\rm (c)}}$ does not
contribute to the entropy but only to the CF.

The diagram (d) in Fig. \ref{fig:entropies} contains a $\rho\rho\pi\pi$ vertex
that comes from Eq. (\ref{4vertex}). This interaction is required by the gauge
invariance of the $\rho$ with the contribution to $m_{\rm el}$ given by
\be \log Z^{\pi\pi}_{{\rm
(d)}}(\mu)=-\frac{3\beta V
g^2}{16}\left[(U_+^\rho+U_-^\rho+2D^\rho)(U_+^\pi+U_-^\pi+2D^\pi)-4D^\rho
D^\pi\right].
\label{logzeight}
\ee 
The upper index for $U,D$ indicates which mass has to be used in the
meson energy $\omega$ of Eqs. (\ref{c_d}) and (\ref{u_v}).

\subsection{Charge conservation}
\label{appendixc3}
In a calculation of CF the conservation of charge is
essential and therefore gauge invariance of the diagrams must be ensured. The
set of diagrams in Fig. \ref{fig:diagrams} 
has been constructed using the Ward identity following the procedure outlined in Appendix
\ref{appendixa}. They ought to be charge conserving by construction.  Nevertheless, it is
desirable to have an explicit proof. The diagrams from Fig. \ref{fig:diagrams} 
represent the heavy
$\rho$ limit of the ones with dynamical $\rho$-mesons in
Fig. \ref{fig:overview_2loop} as shown in Appendix \ref{appendixc}. Therefore, it is 
enough to show charge conservation for the latter.

\begin{figure}
\begin{picture}(450,180)
\put(140,0){ \includegraphics[width=200pt]{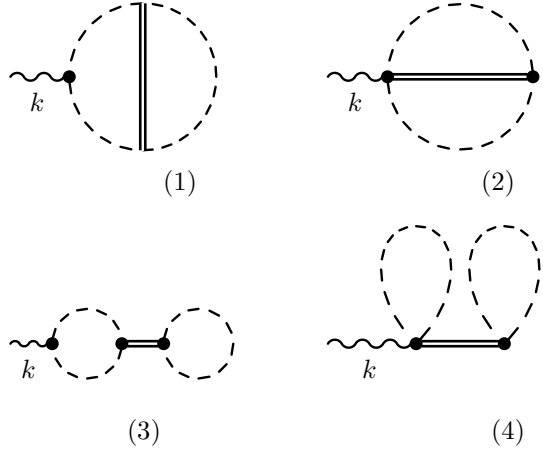}}
\end{picture}
\caption{Elementary
diagrams to be saturated with an additional photon in 
order to construct the selfenergy from Fig. \ref{fig:overview_2loop}.}
\label{fig:body}
\end{figure}

From Ref. \cite{PeSc1995} we utilize the
part of the proof that concerns closed loops. The main statement extracted
from Ref. \cite{PeSc1995} is, adapted to the current situation: Define a
diagram with one external photon at momentum $k$, not necessarily
on-shell. By inserting another photon in all possible ways in the diagram, a set
of new diagrams of photon selfenergy type emerges: For example, the four
diagrams in Fig. \ref{fig:body}
lead to the photon self energies in the two left columns of
Fig. \ref{fig:overview_2loop} plus the (vanishing) diagrams (1c2), (4c), and (5c2) from Fig. \ref{fig:overview_more}, once saturated with an additional photon (we do not allow direct $\gamma\rho\rho$ and $\gamma\gamma\rho\rho$ vertices). The
selfenergy diagrams $\Pi^{\mu\nu}_i$ constructed in this way are charge conserving, and $k_\mu
\sum_i\Pi^{\mu\nu}_i=0$ for the sum over all diagrams. 

For this statement, it has to be shown first that indeed the diagrams from
Fig. \ref{fig:overview_2loop}, including all symmetry and isospin factors,
turn out from the ones of Fig. \ref{fig:body}. This short exercise reveals
that there are two classes of selfenergy diagrams: one comes from
inserting photons in diagrams (1) and (2) of Fig. \ref{fig:body} and the other one from inserting photons in (3) and (4). 
Thus, there are
two separate gauge-invariant classes. 
In a second step, one has to
show the statement from
Ref. \cite{PeSc1995} for the current theory which is different from QED and
richer in vertices of different type:

\vspace*{0.2cm}

\noindent
(I) The $\gamma\pi\pi$ couplings in Fig. \ref{fig:body} can be
transformed into $\gamma\gamma\pi\pi$ couplings by inserting an additional
photon. The $\rho\pi\pi$ vertex can be transformed into a $\gamma\rho\pi\pi$
vertex. These transformations which are a consequence of the momentum
dependence of the vertices are essential for the proof.

\vspace*{0.2cm}

\noindent
(II) For this proof we do not allow direct $\gamma\rho\rho$ and
$\gamma\gamma\rho\rho$ couplings. However, diagrams which include these couplings as in Fig. \ref{fig:overview_more} form a disjoint
gauge class anyway.

\vspace*{0.2cm}

\noindent
(III) The gauge invariance of the diagrams with dynamical $\rho$ in
Fig. \ref{fig:overview_2loop} survives in the heavy $\rho$ limit: According to
Appendix \ref{appendixc}, the amplitudes at a $\rho$-mass of $m_\rho^i$ are multiplied by
$(m_\rho^i/m_\rho)^2$, with $m_\rho$ the physical mass. Then, the limit
$m_\rho^i\to\infty$ is taken and the effective diagrams of
Fig. \ref{fig:diagrams} turn out. The gauge invariance of these diagrams
follows.

\vspace*{0.2cm}

This simple graphical proof demonstrates the charge conservation, and, turning
the argument around, provides a useful tool to ensure that the amplitudes,
including symmetry and isospin factors, have been correctly determined in
Eqs. (\ref{analytic_2_loop},\ref{easyloops}) and Tab. \ref{tab:num_res_eff}.

\section{Structure of the low density expansion}
\label{app:deriva}
For a motivation of Eq. (\ref{b2mu}), we consider the general expression of Eq. (\ref{virialb2}) for the case of two interacting particles. This result is obtained in \cite{Dashen} after carrying out the trace over particle states, i.e. two integrations over the momenta ${\bf p}_1$ and  ${\bf p}_2$ of the interacting particles: 
\be
B_2(\mu=0)&=&-\;\frac{\beta}{2}\int\frac{d^3{\bf p}_1}{(2\pi)^3}\;\frac{1}{2\omega_1'}\int\frac{d^3{\bf p}_2}{(2\pi)^3}\;\frac{1}{2\omega_2'}\;e^{-\beta \omega_1'}\;e^{-\beta \omega_2'}\;T({\bf p}_1,{\bf p}_2)
\label{firststepb2}
\ee
where $\omega_{1,2}'=\sqrt{{\bf p}_{1,2}^2+m_\pi^2}$.
The momenta ${\bf p}_1$ and  ${\bf p}_2$ are defined in the gas rest frame. 
For simplicity, we consider here only a real $T$-Matrix for the interaction and set $\mu=0$. The extension to finite $\mu$ and complex $T$ is straightforward. Note that in the current normalization, $T$ is connected to $T_J^I$ according to $T\to (32 \pi)\; T_J^I$ (see Eq. (\ref{pwaintegral})).
The integrations in Eq. (\ref{firststepb2}) can be rewritten in terms of the momentum of the 2-particle cluster, ${\bf k}={\bf p}_1+{\bf p}_2$, and the relative momentum in the two-particle c.m. frame, ${\bf Q}$, which implies a Lorentz boost along ${\bf k}$. Using the fact that 
\be
\int \frac{d^3 {\bf p}_1}{2\omega_1'}\;\int \frac{d^3 {\bf p}_2}{2\omega_2'}=\int\frac{d^3{\bf Q}\;d^3{\bf k}}{E\sqrt{E^2+{\bf k}^2}}\;,
\label{intid}
\ee
where $E\equiv s^{1/2}$ is the total energy of the pions in the c.m. system, Eq. (\ref{firststepb2}) can be rewritten with a result corresponding to Eq. (\ref{virialb2}). For the moment we ignore the symmetrization operator $A$ in Eq. (\ref{virialb2}) which will be taken care of below. The Lorentz boost of the statistical exponents in Eq. (\ref{firststepb2}) is in this case easy to carry out and leads to the factor $e^{-\beta\sqrt{{\bf k}^2+E^2}}$ by noting that the invariant energy is given by $E^2=s=(\omega_1'+\omega_2')^2-{\bf k}^2$.

Obviously, no quantum statistical information has entered in Eq. (\ref{firststepb2}). However, in Sec. VIIB of Ref. \cite{Dashen} it is shown that the fermionic or bosonic nature of the particles can be (partly) included by summing over exchange diagrams, i.e., permutating the particles. The final result of this procedure is the replacement of the statistical factors $e^{-\beta\omega'}$ in Eq. (\ref{firststepb2}) by Bose-Einstein factors leading to
\be
B_2(\mu=0)&=&-\;\frac{\beta}{2}\int\frac{d^3{\bf p}_1}{(2\pi)^3}\;\frac{1}{2\omega_1'}\int\frac{d^3{\bf p}_2}{(2\pi)^3}\;\frac{1}{2\omega_2'}\;\frac{1}{e^{\beta \omega_1'}-1}\;\frac{1}{e^{\beta \omega_2'}-1}\;T({\bf p}_1,{\bf p}_2),
\label{secondstep}
\ee
which formally have the appearance of Bose-Einstein factors as shown in Sec. VIIB of \cite{Dashen} (see also the example in Sec. VIIC of \cite{Dashen}). As before in the evaluation of Eq. (\ref{firststepb2}), we can use at this point Eq. (\ref{intid}). In order to obtain the final form of Eq. (\ref{b2mu}), finite charge chemical potential and complex $T$-matrix elements, projected over angular momentum, are straightforward introduced. Also, it is more convenient to express the pion scattering in terms of isospin amplitudes. The final result is shown in Eq. (\ref{b2mu}). 

The Lorentz boost into the c.m. frame at velocity ${\bf v}={\bf k}/(E^2+{\bf k}^2)$, implicitly contained in Eq. (\ref{intid}), has also to be carried out for the statistical factors in Eq. (\ref{secondstep}). Unlike in the simple case of Eq. (\ref{firststepb2}), this leads to the slightly more complex expressions shown in Eq. (\ref{boostbose}). Note the boost is not essential but convenient as the scattering amplitude is easily obtained in the two-particle c.m. frame and some integrations can be carried out analytically. 

Formally, exchange diagrams are included in Eq. (\ref{virialb2}) through the symmetrization operator $A$. Note, however, that in the standard form of the virial expansion, Eq. (\ref{b2known}), effects from exchange diagrams are missing. This highlights again the difference between the virial expansion, Eq. (\ref{b2known}), and the low density expansion in Eq. (\ref{b2mu}).

In fact, Eq. (\ref{secondstep}) is not an unfamiliar expression: Let us put $T=\lambda\equiv$const, i.e. using $\phi^4$ theory with ${\cal L}=-\lambda/4!\,\phi^4$, and calculate thermodynamic observables such as the pressure from $B_2$.  Using the same interaction, we can compute the observables also from thermal loops in the imaginary time formalism (see e.g. Sec. \ref{sec:rho_dynamical}) at order $\lambda$. Results are identical. In Sec. \ref{why} this agreement is reconfirmed for more complex interactions. 

The observation of equivalence of the methods from \cite{Dashen} and the thermal loop expansion is, to our best knowledge, novel; although in \cite{Eletsky:1993hv}, using an effective range expansion for the amplitude, a similar equivalence has been found on the level of Eq. (\ref{b2known}), i.e., without including Bose-Einstein statistics through exchange diagrams.

\section{Solutions for the resummations}
\label{appendixd}
An additional technical complication appears in the evaluation of Eq. 
(\ref{Faddeev}) for the
summation (n) when the structure of the vertices between
$\pi^\pm\pi^\pm$-loops or $\pi^0\pi^\pm$-loops is inspected: The interaction
of Eq. (\ref{eff_strong}) leads to a Feynman rule of the form $\left(p^2+q^2+6
pq\right)$ for the vertex between two charged pion loops of momenta $p$ and
$q$, and of the form $\left(p^2+q^2\right)$ between a charged and a
$\pi^0$-loop, always implying the corresponding shift $p^0\to p^0\pm\mu$ ($q^0\to q^0\pm\mu$)
for the inclusion of finite chemical potential. Therefore, the loops can not be factorized easily in the way
Eq. (\ref{Faddeev}) suggests. In order to cast the resummations in a manageable from, we introduce for every term of the sum $\left(p^2+q^2+6
pq\right)$ an entry in an additional index that runs from 1 to
3. The Eq. (\ref{Faddeev}) is then to be read as a matrix equation in its
variables. With the definitions \be W_\pm=\frac{1}{\pi^2}\int\limits_0^\infty
dq \;\omega\;n[\omega\pm\mu], \quad
X_{\pm}=\frac{1}{\pi^2}\int\limits_0^\infty dq\; \omega^2 n[\omega\pm \mu],
\ee additional to the ones of Eq. (\ref{c_d}) and (\ref{u_v}), the 
entries of the Faddeev-like equations (\ref{Faddeev}) can be cast in the form
\be a_0&=&\left(D,\; m_\pi^2 D,\;
0\right),\non
a_\pm&=&\frac{1}{4}\left(U_++U_-,m_\pi^2\left(U_++U_-\right),\sqrt{6}\left(V_--V_+\right)\right)\non
c_0&=&\left( \begin{array}{l} m_\pi^2 D\\ D\\ 0
\end{array}
\right) 
\non 
c_\pm&=&\frac{1}{4}\left( \begin{array}{l}
m_\pi^2\left(U_++U_-\right)\\ U_++U_-\\ \sqrt{6}\left(V_--V_+\right)
\end{array}
\right)\non l_0&=&\left( \begin{array}{lll} C-3D&m_\pi^2(C-5D)&0\\
\frac{1}{m_\pi^2}\;(C-D)&C-3D&0\\ 0&0&0
\end{array}
\right) 
\non 
l_\pm&=&\frac{1}{8}\left( \begin{array}{lll}
W_+-3U_++W_--3U_-&m_\pi^2\left(W_+-5U_++W_--5U_-\right)&\sqrt{6}\left(3V_+-X_+-3V_-+X_-\right)\\
\frac{1}{m_\pi^2}\left(W_+-U_++W_--U_-\right)&W_+-3U_++W_--3U_-&\frac{\sqrt{6}}{m_\pi^2}\left(X_--V_--X_++V_+\right)\\
\frac{\sqrt{6}}{m_\pi^2}\left(X_--V_--X_++V_+\right)&\sqrt{6}\left(3V_+-X_+-3V_-+X_-\right)&6\left(W_++W_-\right)
\end{array}
\right).
\label{Knochenjob}
\ee 
With this extension Eq. (\ref{Faddeev}) is easily solved.
In order to check for bulk errors, one can expand the result in the coupling
constant, and at order $g^2/m_\rho^2$ Eq. (\ref{logmu_eff}) indeed turns out. At order $g^4/m_\rho^4$ the expansion gives the linear chain of
three loops which also emerges from the diagram (r) at that order, and
the results are identical.

The ring diagram (r) from Fig. \ref{fig:resum_schemes} with $N$ ''small'' loops is
given by 
\be &&\log Z_{({\rm r}),N}(\mu)=\non &&\frac{-(-1)^N\beta V}{2 N
\pi^2}\int\limits_0^\infty dp\;p^2\;{\rm Res}\left[\left(\frac{\Pi_\pm
(p^0)}{(p^0+\mu)^2-\omega^2}\right)^N
n[p^0]+\left(\frac{\Pi_\pm(-p^0)}{(p^0-\mu)^2-\omega^2}\right)^N
n[p^0]+\left(\frac{\Pi_0(p^0)}{(p^0)^2-\omega^2}\right)^N n[p^0]\right]
\label{logzring}
\ee 
for $N\ge 3$. The residue is taken for the variable $p^0$ at the poles of
order $N$ in the right $p^0$ half-plane. The tadpole selfenergies $\Pi_\pm$
and $\Pi_0$ for the charged and neutral pion propagator
in Eq. (\ref{logzring}) are given by
\be
\Pi_\pm(p^0)&=&-\frac{g^2}{4m_\rho^2} \left(\left[(p^0+\mu)^2-\omega^2+2m_\pi^2\right]\left[U_++U_-+2D\right]+6\,(p^0+\mu)(V_--V_+)\right),\non
\Pi_0(p^0)&=&-\frac{g^2}{2m_\rho^2}\left((p^0)^2-\omega^2+2m_\pi^2\right)\left(U_++U_-\right)
\ee
which 
is immediately obtained from $a_\pm$ and $a_0$ in Eq. (\ref{Knochenjob}). 

\begin{figure}
\includegraphics[width=4cm]{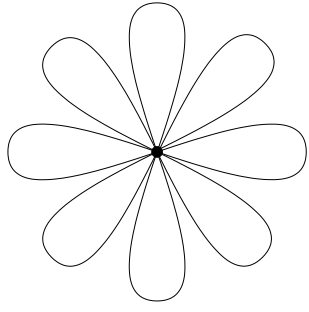}
\caption{Resummation (f) from the expansion of the LO chiral Lagrangian to all orders.}
\label{fig:flowerpower}
\end{figure}
There is an additional possible resummation scheme displayed as (f) in Fig. \ref{fig:flowerpower}. 
The interaction is obtained from the
kinetic term of the LO chiral Lagrangian
Eq. (\ref{chiral}) by expanding it to all orders in the pion fields which
means an exact calculation of the exponentials $U=\exp(i\Phi/f_\pi^2)$ in
Eq. (\ref{chiral}). The mass correction with ${\cal M}$ from
Eq. (\ref{chiral}) is tiny (see Sec. \ref{sec:eff_interaction}) and can be safely neglected. The Lagrangian
for $2n$ fields is then given by ($n\ge 2$): \be {\cal
L}_{2n\pi}^{(2)}=\frac{(-1)^n
4^{n-1}f_\pi^{2(1-n)}}{(2n)!}\;\left((\pi^0)^2+2\pi^+\pi^-\right)^{n-2}
\big((\pi^+\stackrel{\leftrightarrow}{\partial_\mu}\pi^-)^2-2(\pi^-\stackrel{\leftrightarrow}{\partial_\mu}\pi^0)
(\pi^+\stackrel{\leftrightarrow}{\partial^\mu}\pi^0)\big).
\label{toallorders}
\ee 
At $\mu=0$, the grand canonical partition
function from this interaction, summed over all $n$, results in a surprisingly simple expression, 
\be \log
Z_{({\rm f})}(\mu=0)=\frac{\beta V
m_\pi^2}{2}\;\big(f_\pi^2\big(1-e^{-\frac{D}{f_\pi^2}}\big)-D\big)
\label{flowermuzero}
\ee 
with $D$ from Eq. (\ref{c_d}).
This is the special case of the result for finite $\mu$ ($n\geq 2$),
\be \log Z_{({\rm f})}(\mu)&=&\beta
V\sum_{n=2}^\infty\frac{(-1)^n 4^{n-1}f_\pi^{2(1-n)}}{(2n)!}\;
\sum_{k=0}^{n-2} 2^{k-n-1}(-D)^k\sqrt{\pi} (U_++U_-)^{n-k-2}\left(
\begin{array}{c} n-2\\ k\end{array}\right)\non
&\times&\bigg(\;\frac{4D(n-k-2)!\;(m_\pi^2(U_++U_-)^2-(2+k-n)(V_+-V_-)^2)}{\Gamma(-1/2-k)(U_-+U_+)}\non
&+&\frac{\Gamma(n-k)(-m_\pi^2(U_++U_-)(2D+U_++U_-)+(1+k-n)(V_+-V_-)^2)-\Gamma(1-k+n)(V_+-V_-)^2}{\Gamma(1/2-k)}\;\bigg).\non
\label{flowerfinitemu}
\ee 
The sum over $k$ comes from the expansion of the polynomial of order $n-2$
in Eq. (\ref{toallorders}). The possibilities of contracting $2k$ neutral pion
fields have been rewritten,
$\prod_{i=0,k-1}(2k-1-2i)=(-2)^k\Gamma(1/2)/\Gamma(1/2-k)$. 

The structure of the Lagrangian in Eq. (\ref{toallorders}) resembles the 
vertex structure of resummation (t) from Sec. \ref{sec:necklace} with the $\rho\rho\pi\pi$ interaction: 
At any order $n\geq 2$ in the interaction, there are only two
derivative couplings.
Also, the diagrammatic representation of
resummation (t) has the same topology as diagram (f) once the heavy $\rho$ limit
is taken.
Indeed, we observe a close numerical correspondence between the resummations (t) and (f).
Thus, it is interesting to note that the $\rho$ tadpole resummation is well described by an expansion of ${\cal L}_{\pi\pi}^{(2)}$ to all orders.
The resummation (f) is not included in the final numerical results due to these potential double counting problems with (t).

\section{Extension to $\boldsymbol{SU(3)}$}
\label{appendixe}
It is straightforward to extend the study of CF and other thermodynamical observables to $SU(3)$.
Compared to the pion, the other
members of the meson octet have higher masses which simplifies the
selection of relevant processes in a thermal heat bath: 
we regard diagrams which do not contain any
pion as kinematically suppressed.
The contribution to $\log Z$ at $g^2$ then consists 
of diagram (b) in Fig. \ref{fig:entropies} with
one pion line replaced by a kaon and the $\rho$ replaced by the $K^*(892)$.
The  
$\pi KK^*$
interaction follows from Eq. (\ref{34vertex}) in the $SU(3)$ version by extending the representation in Eq. (\ref{su2_fields}) to the full meson and vector meson octet in the standard way
\cite{Marco:1999df,Alvarez-Ruso:2002ib}.  
The result reads 
\be
\log Z_{(b)}^{\pi K}(\mu)&=&-\frac{g^2\beta V}{32}\left[\left(U_+^\pi+U_-^\pi\right)\left(U_+^K+U_-^K+2D^K\right)
+D^\pi\left(U_+^K+U_-^K\right)\right]+\frac{g^2\beta V}{128\pi^4}\left(2m_\pi^2+2m_K^2-m_{K^*}^2\right)\non
&\times&\int\limits_0^\infty dp \int\limits_0^\infty dq\;\frac{pq}{\omega\omega'}\Bigg[\left(
n_+n[\omega'-\mu]+n_-n[\omega'+\mu]+\frac{1}{2}\;n[\omega']\left(n[\omega+\mu]+n[\omega-\mu]\right)
\right)\log_1\non
&&+\left(
n_+n[\omega'+\mu]+n_-n[\omega'-\mu]+\frac{1}{2}\;n[\omega']\left(n[\omega+\mu]+n[\omega-\mu]\right)
\right)\log_2\Bigg]
\label{logzkstarb}
\ee 
where $\omega^2=q^2+m_K^2$, $\omega'^2=p^2+m_\pi^2$, $n_\pm=n[\omega\pm\mu]+n[\omega]$, and the upper index specifies the mass that has to be used in the
definitions of $D$ and $U$ from Eqs. (\ref{c_d}) and
(\ref{u_v}). The expressions $\log_1$ and $\log_2$ are given by 
Eq. (\ref{log1log2}) with the replacement $m_\rho\to m_{K^*}$ and $\omega, \;\omega'$ defined as in Eq. (\ref{logzkstarb}).

Diagram (c) from Fig. \ref{fig:entropies} with $\pi$, $\rho$, and $K$ is possible. 
Also, the $K^* K^*\pi\pi$ term from Eq. (\ref{4vertex}) is present,
shown in Fig. \ref{fig:entropies} (d) with the $\rho$ replaced by a $K^*(892)$. The corresponding contributions are
\be
\log Z_{(c)}^{\pi K}(\mu)&=&-\frac{g^2\beta V}{8m_\rho^2}\left(V_+^\pi-V_-^\pi\right)\left(V_+^K-V_-^K\right),\non
\log Z_{(d)}^{\pi K}(\mu)&=&-\frac{3g^2\beta V}{32}\left[\left(U_+^{K^*}+U_-^{K^*}+2D^{K^*}\right)\left(U_+^\pi+U_-^\pi+D^\pi\right)
-2D^{K^*}D^\pi\right].
\label{cdkstar}
\ee
The electric mass from Eqs. (\ref{logzkstarb}) and (\ref{cdkstar}) is plotted as ''$\pi K$
dynamical'' in Fig. \ref{fig:compare_more}.

The $\pi K$ interaction can be alternatively described 
by the LO chiral Lagrangian from
Eq. (\ref{chiral}) in the $SU(3)$ version (we do not try to construct an effective, point-like, $\pi K$ interaction from $K^*(892)$ exchange as it has been done for $\pi\pi$ via $\rho$ exchange). Using similar arguments as above, the calculation
is reduced to diagram (a) in Fig. \ref{fig:entropies}, with one pion replaced by a kaon. Taking only the kinetic part of Eq. (\ref{chiral}) --- contributions
from the mass term are tiny --- one obtains 
\be \log Z_{(a)}^{\pi K}(\mu)=\frac{-\beta V}{96 f_\pi\; f_K}\;
\left[6\left(V_+^\pi-V_-^\pi\right)\left(V_+^K-V_-^K\right)+\left(m_\pi^2
+m_K^2\right)\left(U_+^K+U_-^K+2D^K\right)
\left(U_+^\pi+U_-^\pi+D^\pi\right)\right]
\label{logmu_eff_K}
\ee 
with $f_K=1.22f_\pi$ taken from chiral perturbation theory \cite{Gasser:1984gg}.
The contribution from Eq. (\ref{logmu_eff_K}) is plotted as ''$\pi K$ contact'' in
Fig. \ref{fig:compare_more} with the dotted line.

In a similar way as in Sec. \ref{sec:relavir}, it is possible to establish a density expansion  for the $\pi K$ interaction that respects
the Bose-Einstein statistics of the asymptotic states in $\pi K$ scattering. Following the same steps as 
in Sec. \ref{sec:relavir}, we obtain, again assuming elastic unitarity (the 50 \% inelasticity in the $\delta_2^{1/2}$ partial wave changes the result only slightly),
\be
B_2^{(\pi K),\;{\rm Bose}}(\mu)&=&\frac{\beta}{4\pi^3}\;\int\limits_{m_\pi+m_K}^\infty dE\int\limits_{-1}^1 dx\int\limits_0^\infty dk\;
\frac{E\;k^2}{\sqrt{E^2+k^2}}\sum_{\ell=0,1,2,\cdots}(2\ell+1)\non
&\times&2\bigg[\delta_\ell^{3/2}\left(n[\omega_\pi+\mu]n[\omega_K+\mu]+n[\omega_\pi-\mu]n[\omega_K-\mu]\right)\non
&&+\frac{1}{3}\left(\delta_\ell^{1/2}+2\delta_\ell^{3/2}\right)\left(n[\omega_\pi]n[\omega_K+\mu]+n[\omega_\pi]n[\omega_K-\mu]\right)\non
&&+\frac{2}{3}\left(\delta_\ell^{1/2}+2\delta_\ell^{3/2}\right)\left(n[\omega_\pi+\mu]n[\omega_K]+n[\omega_\pi-\mu]n[\omega_K]+
n[\omega_\pi]n[\omega_K]\right)\non
&&+\frac{1}{3}\left(2\delta_\ell^{1/2}+
\delta_\ell^{3/2}\right)\left(n[\omega_\pi-\mu]n[\omega_K+\mu]+n[\omega_\pi+\mu]n[\omega_K-\mu]\right)\bigg].
\label{pikvirfull}
\ee
The boosted Bose-Einstein factors are
\be
&&n[\omega_{\pi,K}\pm \mu]=\frac{1}{e^{\beta(\omega_{\pi,K}\pm \mu)}-1},\quad
\omega_{\pi}=\gamma_f\left(E_{\pi}+\frac{k\;Q\;x}{\sqrt{E^2+k^2}}\right),\quad
\omega_{K}=\gamma_f\left(E_{K}-\frac{k\;Q\;x}{\sqrt{E^2+k^2}}\right),
\non
&&\gamma_f=\left(1-\frac{k^2}{E^2+k^2}\right)^{-\frac{1}{2}},
\quad
E_{\pi,K}=\sqrt{Q^2+m_{\pi,K}^2}=\frac{E^2+m_{\pi,K}^2-m_{K,\pi}^2}{2E}
\label{boostbosepik}
\ee
with the c.m. momentum of the particles, $Q=1/(2 E)\sqrt{(E^2-(m_\pi+m_K)^2)(E^2-(m_\pi-m_K)^2)}$. 
For $\mu=0$ and in the Boltzmann limit Eq. (\ref{pikvirfull}) reduces to the virial coefficient
\be
B_2^{(\pi K),\;{\rm Boltz}}(\mu=0)=\frac{1}{2\pi^3}\int\limits_{m_\pi+m_K}^{\infty}dE\;E^2\;K_1(\beta E)\;4
\sum_{\ell=0,1,2,\cdots}(2\ell+1)\left(4\delta_\ell^{3/2}+2\delta_\ell^{1/2}\right)
\label{Boltzmannpik}
\ee
which shows the correct ratio of degeneracy between $\delta_\ell^{3/2}$ and $\delta_\ell^{1/2}$, but is an overall factor
of 4 larger than one would expect --- compare, e.g., to Eq. (\ref{b2known}): Instead of a sum over isospin of the form $\sum_{I,\ell} (2I+1)(2\ell+1)\delta_\ell^I$, the projection 
of charge channels of pions and kaons to the isospin channels leads to $4 \sum_{I,\ell} (2I+1)(2\ell+1)\delta_\ell^I$.
In any case, the result Eq. (\ref{pikvirfull}) for $m_{\rm el}$, using the chiral $\pi K$ interaction at $1/(f_\pi f_K)$, matches exactly the 
thermal loops in Eq. (\ref{logmu_eff_K}). This we have shown in the same way as in Sec. \ref{why} by using Eq. (\ref{deltatcorresp})
and the partial waves $T^{1/2}=(7u-5s-2t)/(12 f_\pi f_K)$ and $T^{3/2}=(2s-t-u)/(6f_\pi f_K)$ (as in Eq. (\ref{logmu_eff_K}), we consider only the kinetic term of ${\cal L}_{\pi K}^{(2)}$). A similar test has been performed by starting from Eq. (\ref{Boltzmannpik}) and using the partial waves from above. From this the pressure has been calculated and results are identical to the pressure obtained from Eq. (\ref{logmu_eff_K}) by taking the Boltzmann limit of the statistical factors $n$ in the definition of $D$, $U$, and $V$. 
Additionally, an independent check for the Lorentz structure of Eq. (\ref{pikvirfull}) has been performed in the same way as in Sec. \ref{why}, this time for a $\phi_1^2\phi_2^2$ interaction of uncharged bosons with different masses $m_{\phi_1}$ and $m_{\phi_2}$.


\end{document}